\documentclass[a4paper,11pt]{article}
\pdfoutput=1 % if your are submitting a pdflatex (i.e. if you have
\usepackage{jcappub} % for details on the use of the package, please
\usepackage[T1]{fontenc} % if needed
\usepackage{epstopdf}

\DeclareMathAlphabet{\mathpzc}{OT1}{pzc}{m}{it}

\title{Charged anisotropic strange stars in Finslerian geometry}

\author[a]{Sourav Roy Chowdhury,}
\author[a]{Debabrata Deb,}
\author[b,c]{Saibal Ray}
\author[d]{Farook Rahaman,}
\author[a]{and B. K. Guha}

\affiliation[a]{Department of Physics, Indian Institute of Engineering Science and Technology,
	Shibpur,\\ Howrah 711103, West Bengal, India}
\affiliation[b]{Department of Physics, Government College of Engineering and Ceramic Technology,\\
	Kolkata 700010, West Bengal, India}
\affiliation[c]{Department of Natural Sciences,
Maulana Abul Kalam Azad University of Technology, Haringhata 741249, West Bengal, India}
\affiliation[d]{Department of Mathematics, Jadavpur University, \\Kolkata 700032, West Bengal, India}

\emailAdd{sourav.rs2016@physics.iiests.ac.in}
\emailAdd{ddeb.rs2016@physics.iiests.ac.in}
\emailAdd{saibal@associates.iucaa.in}
\emailAdd{rahaman@associates.iucaa.in}
\emailAdd{dean.fa@iiests.ac.in}

\abstract{ We investigate a simplified model for the strange stars in the framework of Finslerian spacetime geometry, composed of charged fluid. It is considered that the fluid consisting of three flavor quarks including a small amount of non-interacting electrons to maintain the chemical equilibrium and assumed that the fluid is compressible by nature. To obtain the simplified form of charged strange star we considered constant flag curvature. Based on geometry, we have developed the field equations within the localized charge distribution. We considered that the strange quarks distributed within the stellar system are compiled with the MIT bag model type of equation of state (EOS) and the charge distribution within the system follows a power law. We represent the exterior spacetime by the Finslerian Ressiner-Nordstr{\"o}m space-time. The maximum anisotropic stress is obtained at the surface of the system. Whether the system is in equilibrium or not, has been examined with respect to the Tolman-Oppenheimer-Volkoff (TOV) equation, Herrera cracking concept, different energy conditions and adiabatic index. We obtain that the total charge is of the order of 10$^{20}$ C and the corresponding electric field is of around 10$^{22}$ V/m. The central density and central pressure vary inversely with the charge. Varying the free parameter (charge constant) of the model, we find the generalized mass-radius variation of strange stars and determine the maximum limited mass with the corresponding radius. Furthermore, we also considered the variation of mass and radius against central density respectively.}

\begin{document}
\maketitle
\flushbottom

\section{Introduction}

The theoretical study of the nature and properties of the strange quark stars is an attractive topic of research, not only due to a distinct branch of compact stars, but also the strange matter EOS is worthy of explaining a few astrophysical compact objects. The possibility of strange quark matter (SQM)~\cite{Farhi,Alcock1,Haensel}, made up of equal unconfined up, down and strange quarks may be counted as the basic state of the strong interaction, proposed by Bodmer~\cite{Bodmer} and later on by Witten~\cite{Witten}. Theoretical presence of strange quark star was proposed by Itoh~\cite{Itoh}.

To provide the chemical equilibrate of the strange stars, a small number of electrons should be included in SQM. This small number of electrons (not bounded by any strong interaction) plays a significant role in the constitution of electric dipole layer at the surface. Therefore, a non-negligible electrical energy density comparable to radial pressure has developed a local non-neutral strong electric field within the star, i.e. between electron layer and positively charged core. Strange quark stars in the presence of the strong electric field can be modeled using the Maxwell-Einstein field equations. Strong impacts of this electric filed on the gravitationally bounded system are already shown by the authors~\cite{Alcock2,Usov1,Usov2}. The electric field on the surface region is in the order of $ 10^{19} \sim 10^{20}$ V/m. This surface electric field would be more extreme if the strange matter is made up of color superconducting strange matter~\cite{Alford}.
The strange star demands more charge to be in stable equilibrium in a strong gravitational field. A significant amount of charge can induce an acute electric field. Non-zero charge modulate the structure of the strange star in a different manner: (i) Curvature of space-time, i.e. the metric, (ii) Energy density affiliated to the electric field enriches the total mass of the system, (iii) Coulombian interaction has a finite contribution to the hydrostatic equilibrium of the system, (iv) Charge also contributes to the anisotropic stress. It reduces the amount of anisotropic stress of the system. The charge and mass density contribute finitely to form an equilibrium configuration of the charged fluid.

Within last two decades, various charge distribution has been frequently applied to describe the effects of charge on the interior of strange stars~\cite{Negreiros,Malheiro,Arbanil1}. Polytropic stars with charge density related to the energy density have been studied in~\cite{Ray,Arbanil2}. In their pioneer work, Negreiros  et al.~\cite{Negreiros} investigated the strange star with the Gaussian charge distribution. They found that the charge gradient ($dQ/dr$) is not dependent on the width of charge distribution. This is true for any relativistic stellar object, provided that the distribution is narrowly spread over the system. Several authors worked with the charge distribution in the form of power law $q(r)=Q(r/R)^n$, where $Q$ and $R$ are total charge and radius of the system respectively. Felice et al.~\cite{Felice1,Felice2}, in his works studied the system for n$ \geq $ 3 to make sure that the charge density does not diverge at the origin. Arba{\~n}il and Malheiro, in their work~\cite{Arbanil1} studied for $n=3$ for perfect fluid system and later on, Deb et al.~\cite{Deb1} studied the strange star for the same. Anninos and Rothman~\cite{Anninos} studied the stellar system with a complex type of charge distribution.    

Considering the homogeneous distribution of matter and constant surface charge, the stability of the charged star investigated in~\cite{Stettner} and found that the incompressible star with no charge is less stable than a star with small surface charge and constant energy density. Glazer developed Chandrasekhar's pulsation equation for the charged fluid system~\cite{Glazer1}. Stability of an incompressible fluid star can be increased by introducing charge~\cite{Glazer2}. The hydrostatic equilibrium and collapse of a charged fluid system studied in~\cite{Bekenstein}. An extensive study of the equilibrium dependency and stability on the charge distribution is provided in~\cite{Arbanil1}. They found that the stability of the strange star inversely varies with the total charge due to a certain range of the central energy density, whereas for a range of total mass stability increases with the increment of charge. Also, studied the hydrostatic equilibrium and stability with the radial perturbation for the varying central density, charge and charge-radius ratio.

In this article, we have studied the generalized structure of strange stars in the presence of electric charge, in the framework of Finsler geometry, and study the stability of the system. The reason for choosing Finsler geometry is that its length elements are not bounded by any quadratic restriction and the geometry depends on dynamics along with the position of the system~\cite{Bao}. Extensions of Einstein gravity in Fineslerain geometry is studied in several literature~\cite{Rutz,Vacaru,Pfeifer1}. Nowadays several authors are using Finsler geometry to describe the violation of Lorentzian invariance and anisotropy of the Universe~\cite{Girelli,Gibbons,Chang,Kostelecky1,Kostelecky2,Kouretsis1,Kouretsis2,Li1}. Pfeifer and Wohlfarth~\cite{Pfeifer2} studied the casual structure and the generalized theory of electrodynamics in Finslerian space-time. However, the solutions to the field equations are not discussed. Li \cite{Li2} has obtained a Finslerian Reissner-Nordstr{\"o}m solution for vacuum space. He also found the eigenfunction of the
Finslerian Laplacian operator. The generalized form of the Maxwell-Einstein Field equation in the following geometry is obtained  from the geodesic equation of motion in the geometry. We have considered simplified MIT bag model EOS and the charge distribution in the form as assumed in~\cite{Arbanil1} (power law). For simplicity, we constrained ourselves in constant flag curvature. Akbar-Zahed~\cite{Akbar} already discussed the helpfulness of considering constant flag curvature and generality remain same. We obtained the exact solution of the Maxwell-Einstein Field equation. As a consequence, obtained maximum enclosed charge and field on the surface. The generalized variation of mass-radius relation for the strange stars due to a definite value of bag has been enumerated. We found a range of total mass, radius and charge respective of central density along with the theoretical bounds. The explicit study of the system are shown graphically and in tabular format for $\overline{Ric} =1.2$ and bag value $83~Mev/fm^3 $, which is within the well accepted range. 

Our paper is structured as follows:~Definition and formation of field equation of Finsler geometry are outlined in section \ref{sec:formation}. $Adhoc$ relations are stated in section \ref{sec:adhoc}. We provided the formalism of basic stellar equations in section \ref{sec:solu}, we have presented the solution of the Maxwell-Einstein field equations. Physical acceptability and stability of the stellar system are verified in section \ref{sec:physical}  by studying mass-radius relation, energy conditions verification, the stability of the stellar model and compactification. Finally, the conclusion of our study with a discussion is provided in section \ref{sec:outlook}.

\section{BASIC STELLAR EQUATIONS}
\label{sec:formation}
We briefly present fundamental geometrical concepts from the
theory of Finsler spaces and generates the respective field equations, as well as discuss about the EOS of the stellar system.

\subsection{Basic Formalism}

We consider on a manifold $ \mathpzc{M}$, the Finsler metric is $F$. In standard coordinate notation, $F = F( x, \dot{x})$ is the function of $(x^\mu, y^\mu)$ in  $\in T \mathpzc{M}$.

The equation of Geodesic of the Finsler metric $(F)$ can be written as follows:
\begin{equation}
\frac{d^2x^\mu}{d \tau^2}+2G^\mu(x, \dot{x}) =0, \label{1},
\end{equation}
where the geodesic spray is given by,
\begin{equation}
G^\mu = \frac{1}{4} g^{\mu \nu} \left( \frac{\partial ^2F^2}{\partial x^\lambda \partial y^\nu} y^\lambda -\frac{\partial F^2}{\partial x^\nu} \right). \label{2}
\end{equation}

The metric structure coefficient is given by
\begin{equation*}
g_{\mu \nu} = \frac{\partial }{\partial y^\mu} \frac{\partial }{ \partial y^\nu} \left( \frac{1}{2} F^2 \right),
\end{equation*}
with $(g ^{\mu \nu} )$ = $ (g_{\mu \nu})^{-1} $.

Let, the Finsler structure be
\begin{equation}
F^2= e^{\lambda(r)} y^t y^t - e^{\nu(r)} y^r y^r -r^2 \overline{F}^2(\theta, \phi, y^\theta, y^\phi ). \label{3}
\end{equation}

 Using eq. (\ref{1}), the Finsler metric potential can be defined as
\begin{eqnarray}
&g_{\mu \nu} = diag(e^{\lambda(r)},- e^{\nu(r)}, -r^2\overline{g}_{ij}).  \label{4} 
%&g^{\mu \nu} = diag(e^{-\lambda(r)},- e^{-\nu(r)}, -r^2\overline{g}^{ij}) \nonumber
\end{eqnarray}

The respective geodesics sprays of the system are as follows:
\begin{eqnarray}
& G^t = \frac{1}{2}\lambda' y^t y^r, \label{5} \\
& G^r = \frac{1}{4}  \left( \nu' y^r y^r +\lambda' e^{\lambda-\nu} y^t y^t -2re^{-\nu}\overline{F}^2  \right),  \label{6} \\
& G^\theta = \frac{1}{r} y^\theta y^r +\overline{G}^\theta, \label{7}\\
& G^\phi = \frac{1}{r} y^\theta y^r +\overline{G}^\phi.  \label{8}
\end{eqnarray}

In Finsler geometry, Ricci tensor is introduced by Akbar-Zadeh~\cite{Akbar}, given by
\begin{equation}
Ric_{\mu\nu}= \frac{\partial^2}{\partial y^\mu \partial y^\nu}\left( \frac{1}{2}F^2 Ric \right). \label{9}
\end{equation}

Ricci scalar in the Finsler geometry is,  
\begin{equation}
	Ric =R^\mu_\mu=\frac{1}{F^2} \left[ 2 \frac{\partial G^\mu}{\partial x^\mu}- y^\lambda \frac{\partial^2 G^\mu}{\partial x^\lambda \partial y^\mu} +2 G^\lambda \frac{\partial^2 G^\mu}{\partial y^\lambda \partial y^\mu} - \frac{\partial G^\mu}{\partial y^\lambda}\frac{\partial G^\lambda}{\partial y^\mu} \right], \label{10}
\end{equation}
where, $R^\mu_\mu  $ is insensitive to connections, it only depends on Finsler structure.

 With the help of eqs. (\ref{2}), (\ref{3}) and (\ref{10}), we obtain,
\begin{eqnarray}
& & F^2 Ric= \left[ \frac{\lambda''}{2}+\frac{\lambda'^2}{4}-\frac{\lambda' \nu'}{4} +\frac{\lambda'}{r} \right] e^{(\lambda-\nu)} y^t y^t + \left[- \frac{\lambda''}{2}-\frac{\lambda'^2}{4} +\frac{\lambda' \nu'}{4} +\frac{\nu'}{r} \right]y^r y^r \nonumber \\
& & \hspace{5cm} +\left(\overline{Ric}-e^{-\nu} +\frac{r \nu' e^{-\nu}}{2}-\frac{r \lambda' e^{-\nu}}{2}\right).\label{11}
\end{eqnarray}

The scalar curvature can define as $S = g^{\mu \nu} Ric_{\mu \nu}$.

 Therefore, in Finsler geometry, the modified form of Einstein tensor reads,
\begin{eqnarray}
G^{\mu }_{\nu} = g^{\mu \nu}Ric_{\mu \nu}-\frac{1}{2} S, \label{12}
\end{eqnarray}
 with the consideration of the flag curvature reads  $ \overline{F}^2 =y^\theta y^\theta + f(\theta) y^\phi y^\phi $.\\

The explicit form of $S$ for the Finsler structure eq. \ref{3} as follows:
\begin{equation}
S = e^{-\nu}\left[ \lambda''+\frac{\lambda'^2}{2}- \frac{\lambda' \nu'}{2} +\frac{\lambda'-\nu'}{r} \right]  -\frac{2}{r^2} \left[\overline{Ric}-e^{-\nu} +\frac{r  e^{-\nu}}{2} (\nu'- \lambda')\right]. \label{13}
\end{equation}

The energy momentum tensor for the anisotropic fluid distribution and electromagnetic field within the system,  has the following forms respectively,
\begin{eqnarray}\label{14}
& & T^{i}_{j}= (\rho +p_t)u^{i}u_{j} - p_t \delta^{i}_{j} - (p_t-p_r)v^{i}v_{j},\\ \label{15}
& & E^{i}_{j}= \frac{1}{4 \pi}(-\mathcal{F}^{im}\mathcal{F}_{jm} + \delta^i_j\mathcal{F}^{mn}\mathcal{F}_{mn}),
\end{eqnarray}
where $\rho$, $p_r$ and $p_t$ represent the energy density, radial and tangential pressures, respectively.Here, $ u_{i} $ and $ v_{i} $ represent four-velocity and radial four-vector, respectively. $\mathcal{F}^i_j$ is the anti-symmetric electromagnetic field tensor.

Hence, in the Finsler geometry the Einstein-Maxwell field equations is
\begin{equation}
G^\mu_\nu= 8 \pi_F(T^{i}_{j} +E^{i}_{j}) =8 \pi_F T^\mu_\nu, \label{16}
\end{equation}
where, we consider the geometrized unit, i.e., $G=1=c$.

The covariant divergence of the stress-energy tensor is
\begin{equation} 
T^\mu_{~\nu;\mu} = 0.
\end{equation}

The effective energy-momentum tensor for the locally anisotropic charged fluid distribution  can be written as,

\begin{eqnarray}
	T^\mu_\nu= \left(
	\begin{tabular}{cccc}
	$\rho+\frac{q^2}{8\pi r^4}$ & 0 & 0 & 0 \\
	0 & $-(p_r-\frac{q^2}{8\pi r^4})$ & 0 & 0 \\
	0 & 0 & $-(p_t+\frac{q^2}{8\pi r^4})$ & 0\\
	0 & 0 & 0 & $-(p_t+\frac{q^2}{8\pi r^4})$ \\
	\end{tabular}
	\right),
\end{eqnarray}
where, the electric charge ($\mathbf{q}$) and the corresponding field ($\mathbf{E}$) is related as $ \frac{E^2}{8 \, \pi} = \frac{q^2}{8 \, \pi \,r^4} $.

So, the components of the Einstein-Maxwell field equations are given by
\begin{eqnarray}
\frac{\nu' e^{-\nu}}{r} - \frac{ e^{-\nu}}{r^2} +\frac{\overline{Ric}}{r^2} &= 8 \pi_F \rho +E^2, \label{18}\\
\frac{\lambda' e^{-\nu}}{r} + \frac{ e^{-\nu}}{r^2} - \frac{\overline{Ric}}{r^2} &= 8 \pi_F p_r -E^2, \label{19}\\
e^{-\nu}\left[ \frac{\lambda''}{2}+\frac{\lambda'^2}{4}-\frac{\lambda' \nu'}{4} +\frac{\lambda'-\nu'}{2r} \right] &= 8 \pi_F p_t +E^2. \label{20} 
% e^{-\nu}\left[ \frac{\lambda''}{2}+\frac{\lambda'^2}{4}-\frac{\lambda' \nu'}{4} +\frac{\lambda'-\nu'}{2r} \right] &= 8 \pi_F p_t +E^2\label{21} 
\end{eqnarray} 

The effected gravitational mass equation due the spherically symmetric charged compact stellar object is defined by
\begin{equation}
\frac{dm}{dr}=4\, \pi\,r^2 \rho + \frac{q}{r}\frac{dq}{dr}.
\end{equation} 

\subsection{Junction Condition}
The stellar structure extended from centre towards the surface of the system within the following constrains
\[
m(0) = 0, ~~q(0)=0,~~ \rho(0) = \rho_{c}~~and~~p_r(0)=p_c.
\]

The interior spacetime of the stellar system should be matched smoothly at the boundary with the exterior spacetime. The Finslerian Ressiner-Nordstr\"om metric to represent the exterior spacetime of the following form \cite{Li2}
\begin{equation}
F^2= f_{RN} y^t y^t - f_{RN} ^{-1} y^r y^r -r^2 \overline{F}^2(\theta, \phi, y^\theta, y^\phi ), \label{23}
\end{equation}
where $f_{RN}=\left(C-\frac{2GM}{R}+\frac{GQ^2}{r^2}\right)$ with $M$ and $Q$ being the total mass and charge of the system, respectively, and $C$ is a constant.

\section{Charge distribution, Density profile and Equation of state} \label{sec:adhoc}

To study the effects of charge on the anisotropic strange stars, the charge distribution definitely should have a form. Considered the charge distribution is in the form of power law following Felice et al.~\cite{Felice1} as $ q(r)= Q (r/R)^n $. For the simplicity, we assume that $n=3$ as follows,

\begin{equation}
q(r)=Q\left(\frac{r}{R}\right)^3 = \alpha r^3, \label{24}
\end{equation}
where $Q$ is the total charge and $R$ is the radius of the system, respectively. $ \alpha \left(=\frac{Q}{R^3}\right)$ is a constant (charge constant). 

We consider that the energy density of the fluid inside the strange stars maintaining the form  \cite{Mak} 
\begin{equation}
\rho(r)=\rho_c\left[1-\left(1-\frac{\rho_0}{\rho_c}\right)\frac{r^{2}}{R^{2}}\right],\label{25}
\end{equation}
where $\rho_c$ and $\rho_0$ are the central and surface density, respectively.

The SQM within the system is described by the phenomenal MIT bag model. We assumed that the quarks are massless and non-interacting (Here we considered up($u~$), down($d~$) and strange ($s~$) quarks).
The corrected form of pressure can be defined as after introduction of $ adhoc $ bag function (B) as follows,
\begin{eqnarray}\label{MITEOS1}
p_r = \sum_{f=u,d,s} p^f-B,
\end{eqnarray}
where $p^f$ is the pressure of each type of quarks, i.e. u, d and s. The corresponding corrected energy density is as follow,
\begin{eqnarray}\label{MITEOS2}
\rho= \sum_{f=u,d,s} \rho^f+B,
\end{eqnarray}

Now, substituting the relation between The pressure and energy density due to the each quark flavor given by $p^f=\frac{1}{3}\rho^f$ and eq.~\eqref{MITEOS2} into eq.~\eqref{MITEOS1} we have the final form of the MIT bag EOS as follows
\begin{equation}
p_r=\frac{1}{3}(\rho-4B). \label{26}
\end{equation}

The radial pressure ($ p_r $) vanishes on the surface. Therefore, we can consider the surface density ($ \rho_0 $) as 4B. Following that the eqs. (\ref{26}) can be rewritten as,
\begin{equation}
p_r=\frac{1}{3}(\rho-\mathbf{\rho_0}).
\end{equation}

%%%%%%%%%%%%%%%%%%%%%%%%%%%%%%%%%%%%%%%%%%%%%%%%%%%%%%%%%%%%%%%%%%%%%%%%%%%%%%%%%%%%%%%%%%%%%%%%%%%%%%%%

\begin{figure}\centering
	\includegraphics[scale=0.3]{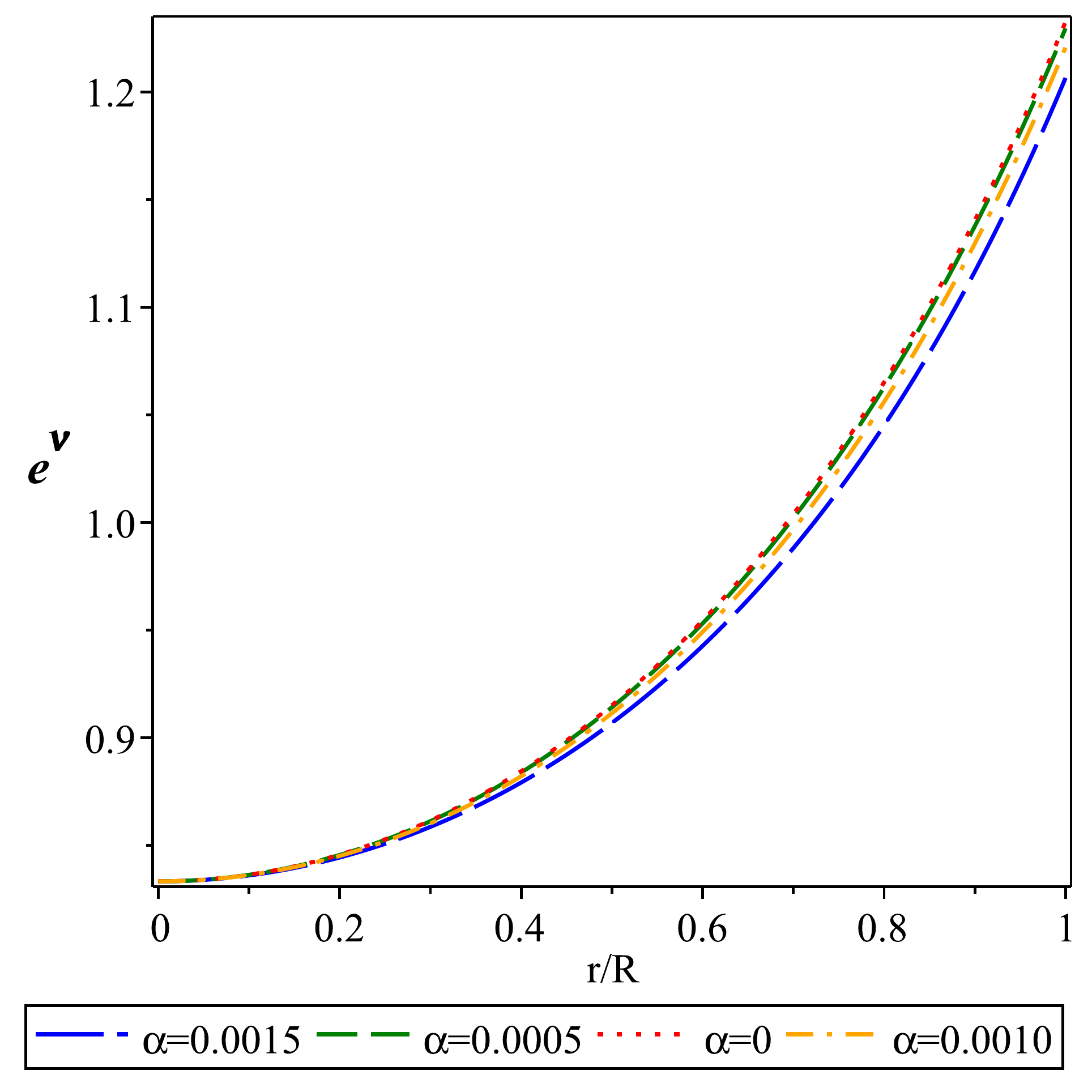}
	\includegraphics[scale=0.3]{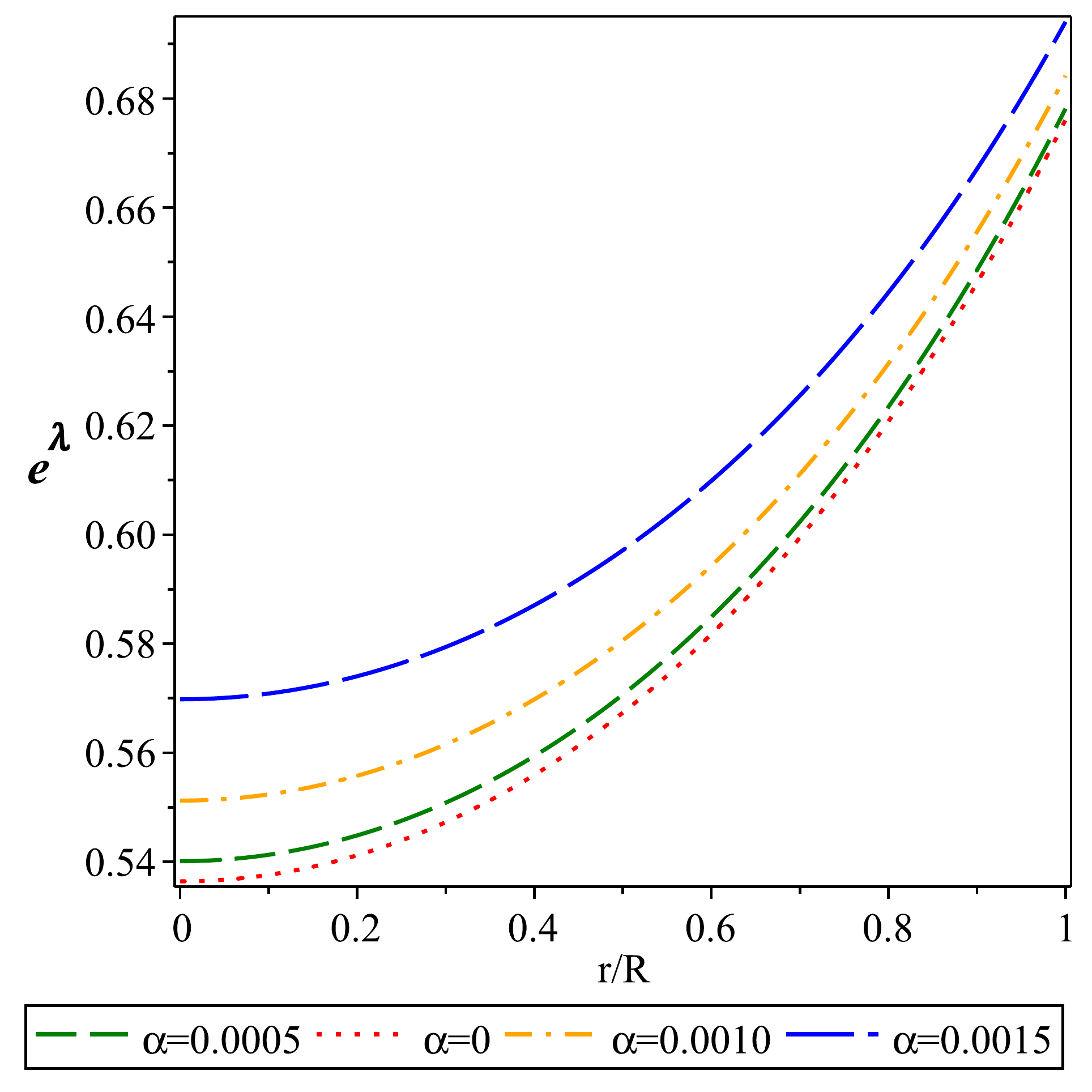}
	\caption{Variation of i) $e^{\nu(r)}$ (left panel) and ii) $e^{\lambda(r)}$ (right panel) as a function of the fractional radial coordinate $r/R$ for the $LMC ~X-4$. Here and in what follows the bag constant $B_g = 83 MeV/fm^3$ and $\overline{Ric} =1.2$.} \label{fn}
\end{figure}

%%%%%%%%%%%%%%%%%%%%%%%%%%%%%%%%%%%%%%%%%%%%%%%%%%%%%%%%%%%%%%%%%%%%%%%%%%%%%%%%%%%%%%%%%%%%%%%%%%%%%%%%

%%%%%%%%%%%%%%%%%%%%%%%%%%%%%%%%%%%%%%%%%%%%%%%%%%%%%%%%%%%%%%%%%%%%%%%%%%%%%%%%%%%%%%%%%%%%%%%%%%%%%%%%

\begin{figure}\centering
	\includegraphics[scale=0.255]{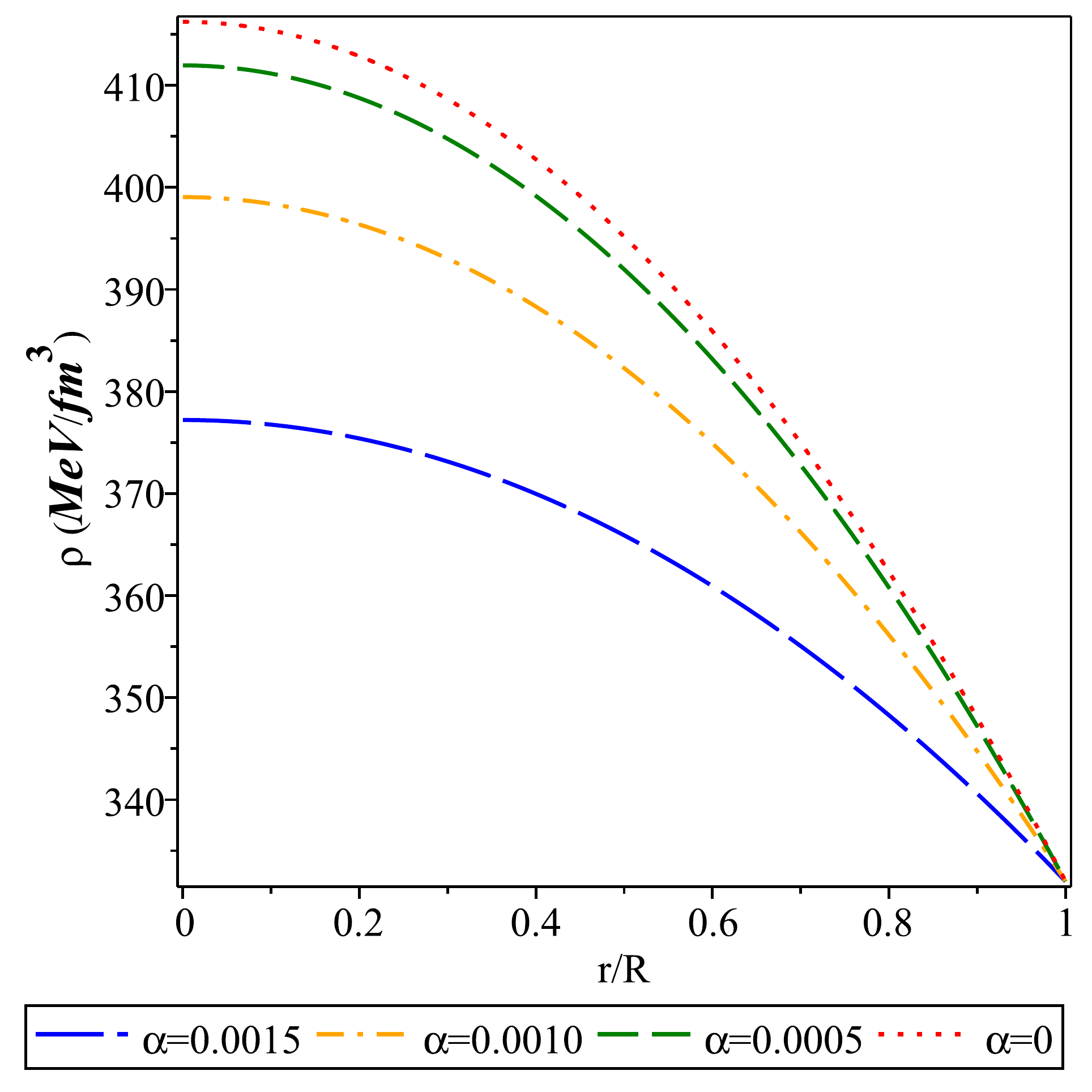}
	\includegraphics[scale=0.255]{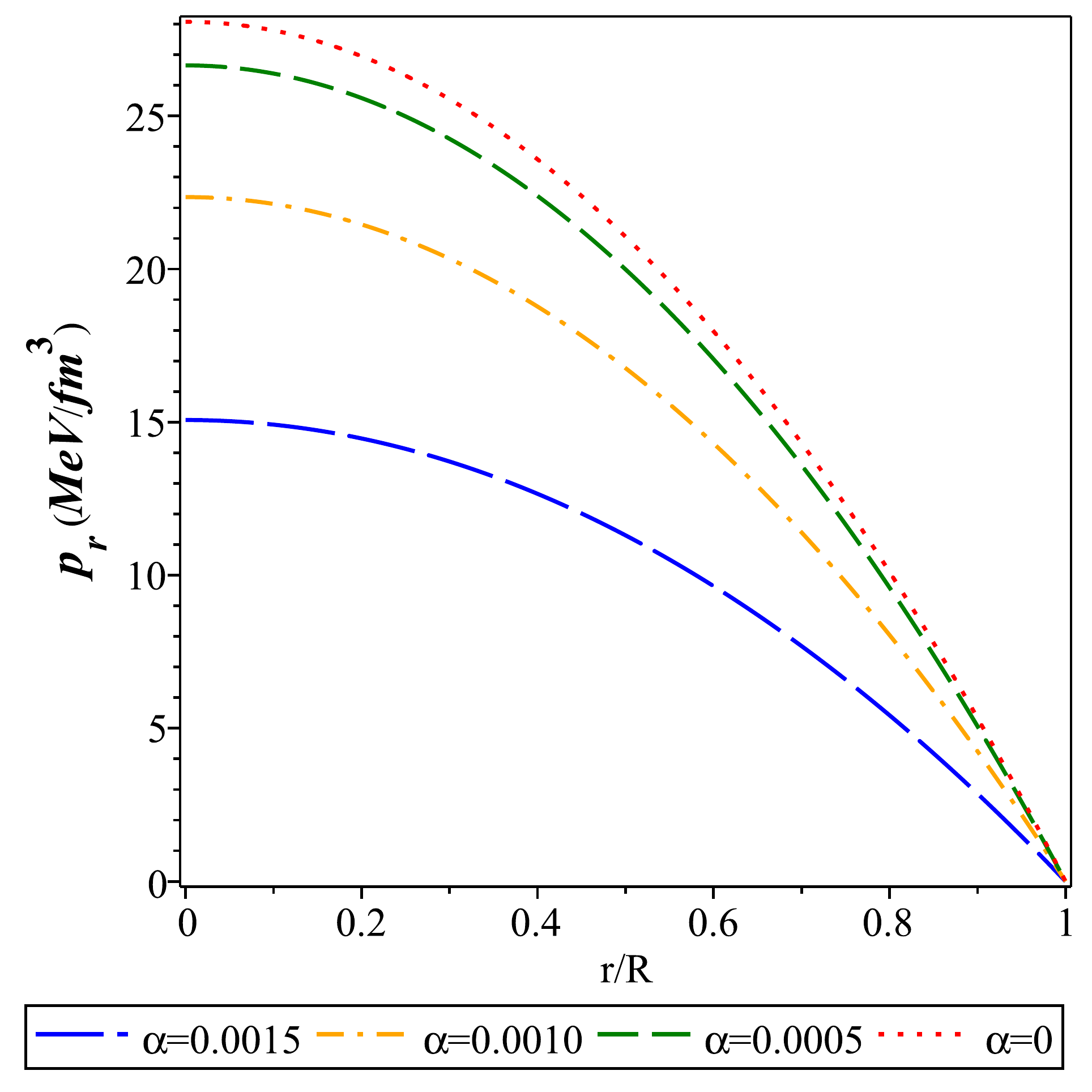}
	\includegraphics[scale=0.255]{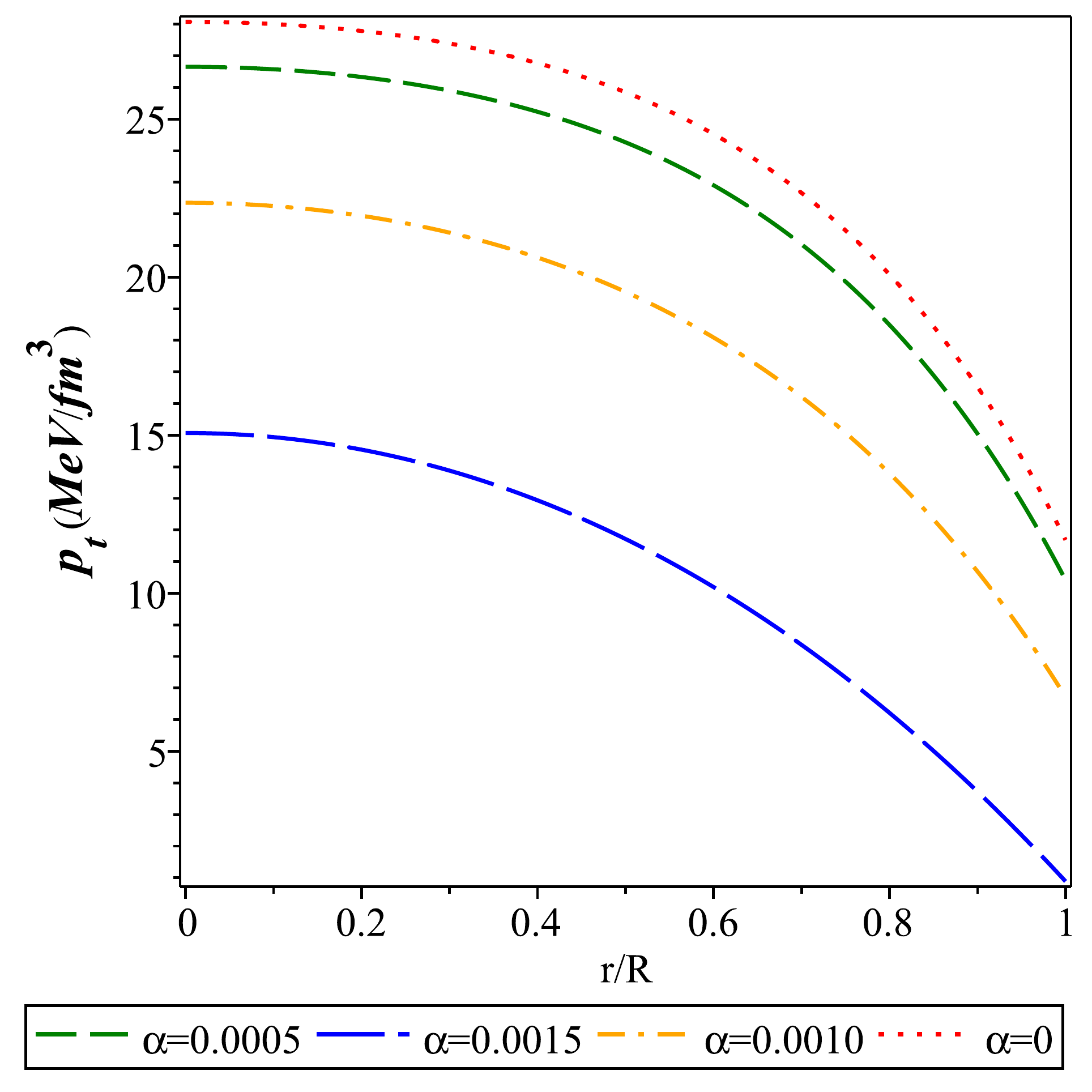}
	\caption{Variation of i) $\rho$ {(left panel)}, ii) $p_r$ {(middle panel)} and iii) $p_t$ {(right panel)} as a function of the fractional radial coordinate $r/R$ for the $LMC ~X-4$.} \label{frho}
\end{figure}

%%%%%%%%%%%%%%%%%%%%%%%%%%%%%%%%%%%%%%%%%%%%%%%%%%%%%%%%%%%%%%%%%%%%%%%%%%%%%%%%%%%%%%%%%%%%%%%%%%%%%%%%

\section{Solution of the Einstein-Maxwell Field Equation}\label{sec:solu}

We obtained the following expression for the gravitational potentials ($\nu$ and $\lambda$), density, radial and tangential pressures, respectively, by solving the eqs.~\eqref{18}-\eqref{20} with the help of eqs.~\eqref{24}-\eqref{26}, given by
\begin{eqnarray}
& & \nu \left( r \right) =-\ln  \left( {\frac { \nu_1 {r}^{4}+ \nu_2 {r}^{2}+A{R}^{5}}{{R}^{5}}} \right), \label{27}\\
& & \lambda (r) =-\frac {32}{\lambda_{{1}}\nu_{1}} \Bigg[\lambda_{1}\,{\rm arctanh} \left({\frac {{R}^{5}{\alpha}^{2}+16\,B\pi\,{R}^{3}-M}{\lambda_{{2}}}}\right) + \lambda_{{2}} \left(\lambda_{3}-\lambda_{4} \right)\ln \lambda_{5}\nonumber\\
& & \hspace{2.5cm} +\lambda_{1}\,{\rm arctanh} \left({\frac {2 \nu_{1}r^2+2\,\nu_{2}}{\lambda_{{2}}{R}^{2}}}\right)+\Big\{\lambda_{4}\,\ln  \left(  \left( \nu_{1}r^2+\nu_{2} \right) {r}^{2}+\overline{{\it Ric}} {R}^{5} \right)\nonumber\\
& & \hspace{2.5cm} -\lambda_{3}\,\ln  \overline{{\it Ric}}  +\frac{5}{2}\left( {\frac {3\,{R}^{5}{\alpha}^{2}}{40}}+B\pi\,{R}^{3}-\frac{3}{16}\,M \right) \,\ln R   \Big\} \lambda_{{2}} \Bigg], \label{28}
\end{eqnarray}
\begin{eqnarray}
& & \rho={\frac {9\,{R}^{5}{\alpha}^{2}{r}^{2}+80\,B\pi\,{R}^{3}{r}^{2} -15\,M{r}^{2}-3\,\nu_2}{8 \pi {R}^{5}}},\label{29}\\
& &	p_{{r}}=-\frac{1}{24}\,{\frac { \left( 9\,{R}^{5}{\alpha}^{2}+80\,B\pi\,{R}^{3}-15\,M \right)  \left( R+r \right)  \left( R-r \right) }{\pi {R}^{5}}}, \label{30}\\
& & p_{{t}}= -\frac {1}{72\,\pi \,(\left( \nu_1 r^2 +\nu_2 \right) r^2 +\overline{Ric} R^5) {R}^{5}}\Big\{ -81\,{R}^{14}{\alpha}^{4}{r}^{2}-54\,{R}^{10}{\alpha}^{4}{r}^{6}-1008\,B\pi\,{R}^{12}{\alpha}^{2}{r}^{2}\nonumber\\
& & \hspace{2.5cm} +1632\,B\pi\,{R}^{10}{\alpha}^{2}{r}^{4} -720\,B\pi\,{R}^{8}{\alpha}^{2} {r}^{6}-3328\,{B}^{2}{\pi}^{2}{R}^{10}{r}^{2}+5888\,{B}^{2}{\pi}^{2}{R}^{8}{r}^{4}\nonumber\\
& & \hspace{2.5cm} -2560\,{B}^{2}{\pi}^{2}{R}^{6}{r}^{6}+1680\,BM\pi\,{R}^{7}{r}^{2} -2304\,BM\pi\,{R}^{5}{r}^{4}+270\,M{R}^{9}{\alpha}^{2}{r}^{2}\nonumber\\
& & \hspace{2.5cm} -360\,M{R}^{7}{\alpha}^{2}{r}^{4}+240\,\overline{{\it Ric}}B\pi\, {R}^{10}-480\,\overline{{\it Ric}}B\pi\,{R}^{8}{r}^{2} +135\,M{R}^{5}{\alpha}^{2}{r}^{6}\nonumber\\
& & \hspace{2.5cm} +27\,\overline{{\it Ric}}{R}^{12}{\alpha}^{2}-27\,\overline{{\it Ric}}{R}^{10}{\alpha}^{2}{r}^{2}+960\,BM\pi\,{R}^{3}{r}^{6}-45\,\overline{{\it Ric}}M{R}^{7}\nonumber\\
& & \hspace{2.5cm} +90\,\overline{{\it Ric}}M{R}^{5}{r}^{2}-225\,{M}^{2}{R}^{4}{r}^{2}+225\,{M}^{2}{R}^{2}{r}^{4}-90\,{M}^{2}{r}^{6}+135\,{R}^{12}{\alpha}^{4}{r}^{4}\Big\},\nonumber \\ \label{31} 
\end{eqnarray}
where $\nu_1, \nu_2,\lambda_{1},\lambda_{2},\lambda_{3},\lambda_{4}$ and $\lambda_{5}$ are constants and their expressions are provided at the {\it Appendix}.

{The variation of the gravitational potentials, viz. $\textrm{e}^\nu$ and $\textrm{e}^\lambda$ as a function of the fractional radial coordinates (r/R) at the interior of the stellar system are exhibited in the left and right panel of figure \ref{fn}, respectively. The variations of the physical quantities like $\rho$, $p_r$ and $p_t$ are shown in the left, middle and right panel of figure \ref{frho}, respectively.}

The anisotropic stress ($ \Delta $) of a stellar system can be expressed as additional tangential pressure over radial direction, i.e. $p_t-p_r$, is given by
\begin{eqnarray}
& & \Delta (r)= \frac{10}{3\pi \,(\left( \nu_1 r^2 +\nu_2 \right) r^2 +\overline{Ric} R^5) {R}^{5}} {r}^{2} \Bigg[ {\frac {27}{40}}\,{R}^{14}{\alpha}^{4}-\frac{9}{4}\,M{R}^{9}{\alpha}^{2}+\Big( {\frac {9\,{r}^{4}{\alpha}^{4}}{20}}-{\frac {77\,B{r}^{2}\pi\,{\alpha}^{2}}{5}}\nonumber \\
& & \hspace{1.3cm} +{\frac {448\,{B}^{2}{\pi}^{2}}{15}} \Big) {R}^{10} +{\frac {15}{8}}\,{M}^{2}{R}^{4} -15\,M\left( B\pi-{\frac {89\,{r}^{2}{\alpha}^{2}}{400}} \right) {R}^{7}+ \left(9\,B\pi\,{\alpha}^{2} -{\frac 
{9}{8}}\,{r}^{2}{\alpha}^{4} \right) {R}^{12}\nonumber \\
& & \hspace{1.3cm} +B\pi\, \left( {\frac {34\,{\alpha}^{2}{r}^{4}}{5}}-{\frac {848\,B\pi\,{r}^{2}}{15}}+\overline{Ric} \right) {R}^{8}-10\,BM\pi\,{R}^{3}{r}^{4}+{\frac {80\,{B}^{2}{\pi}^{2}{R}^{6}{r}^{4}}{3}}\nonumber \\
& & \hspace{1.3cm} -\frac{3}{16}\,M \left( {\frac {34\,{\alpha}^{2}{r}^{4}}{5}}-{\frac {1888\,B\pi\,{r}^{2}}{15}}+\overline{Ric}\right) {R}^{5} -{\frac {39\,{M}^{2}{R}^{2}{r}^{2}}{16}}+{\frac {15\,{M}^{2}{r}^{4}}{16}} \Bigg].\label{32}
\end{eqnarray} 

\begin{figure}\centering
	\includegraphics[scale=0.3]{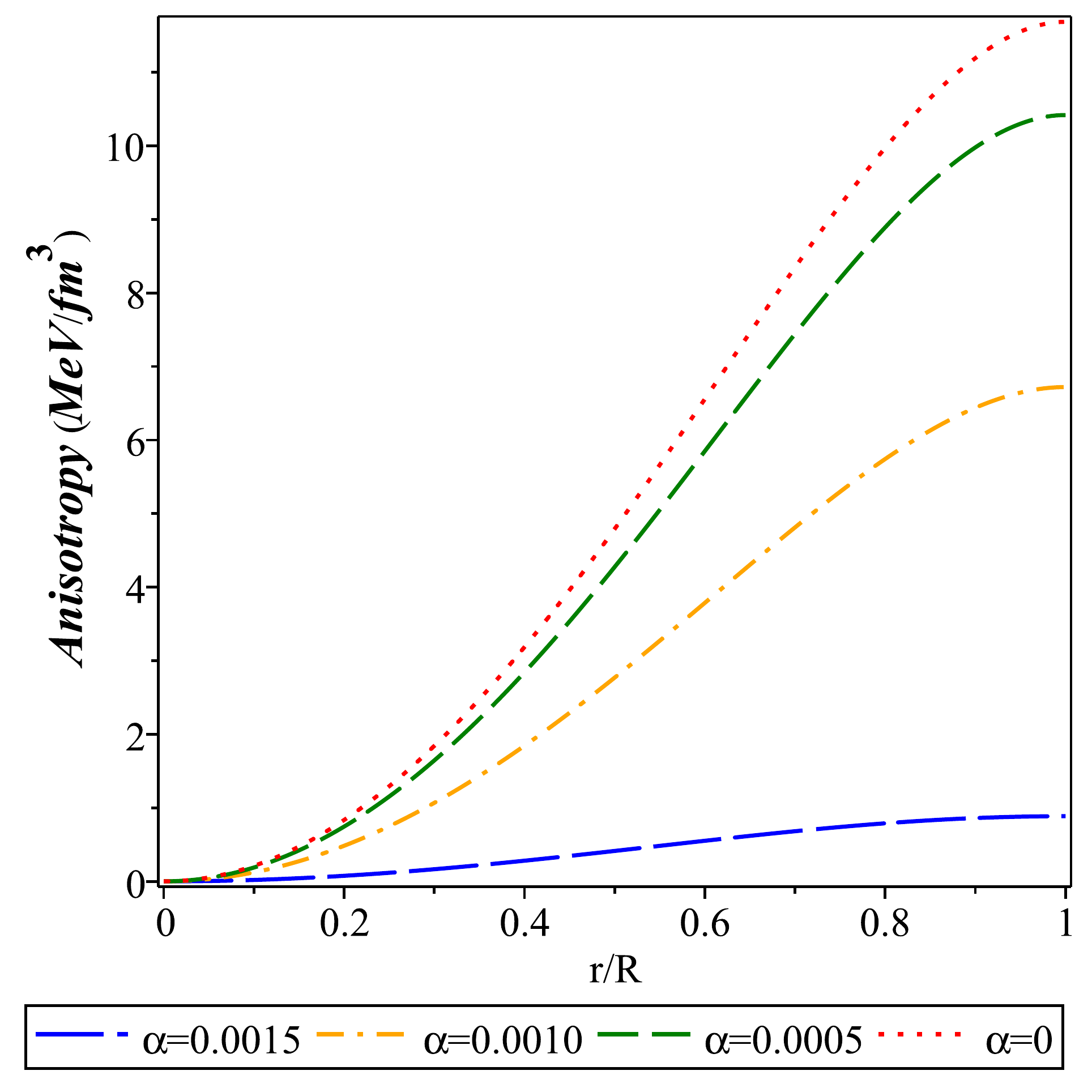}
	\caption{Variation of anisotropic stress $(\Delta)$ as a function of the fractional radial coordinate $r/R$ for the $LMC ~X-4$.} \label{fani}
\end{figure}

 The variation of the anisotropic stress of the stellar system as a function of fractional radial coordinates ($r/R$) is shown in figure \ref{fani}.

%%%%%%%%%%%%%%%%%%%%%%%%%%%%%%%%%%%%%%%%%%%%%%%%%%%%%%%%%%%%%%%%%%%%%%%%%%%%%%%%%%%%%%%%%%%%%%%%%%%%%%%%

%%%%%%%%%%%%%%%%%%%%%%%%%%%%%%%%%%%%%%%%%%%%%%%%%%%%%%%%%%%%%%%%%%%%%%%%%%%%%%%%%%%%%%%%%%%%%%%%%%%%%%%%

\begin{figure}\centering
	\includegraphics[scale=0.3]{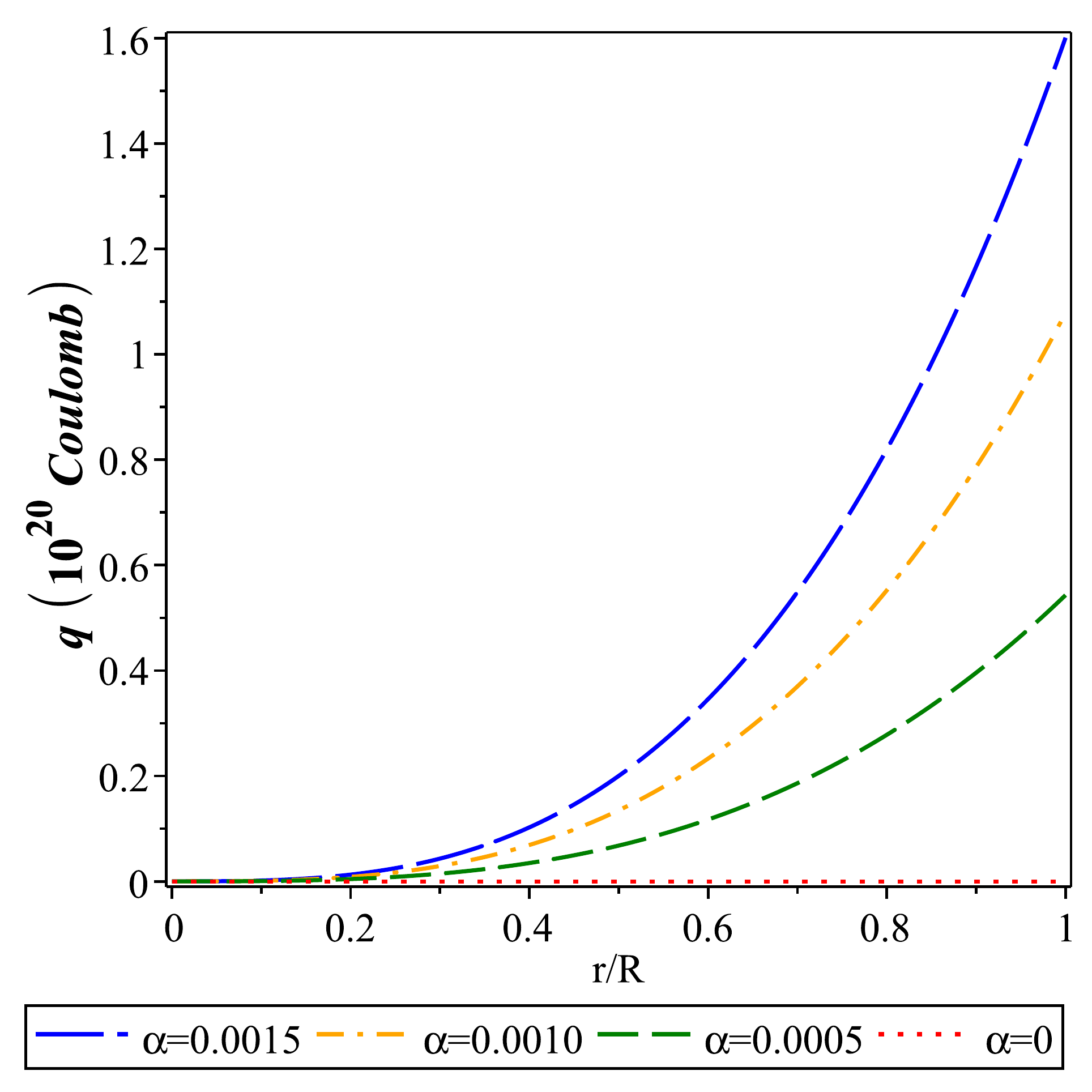}
	\includegraphics[scale=0.3]{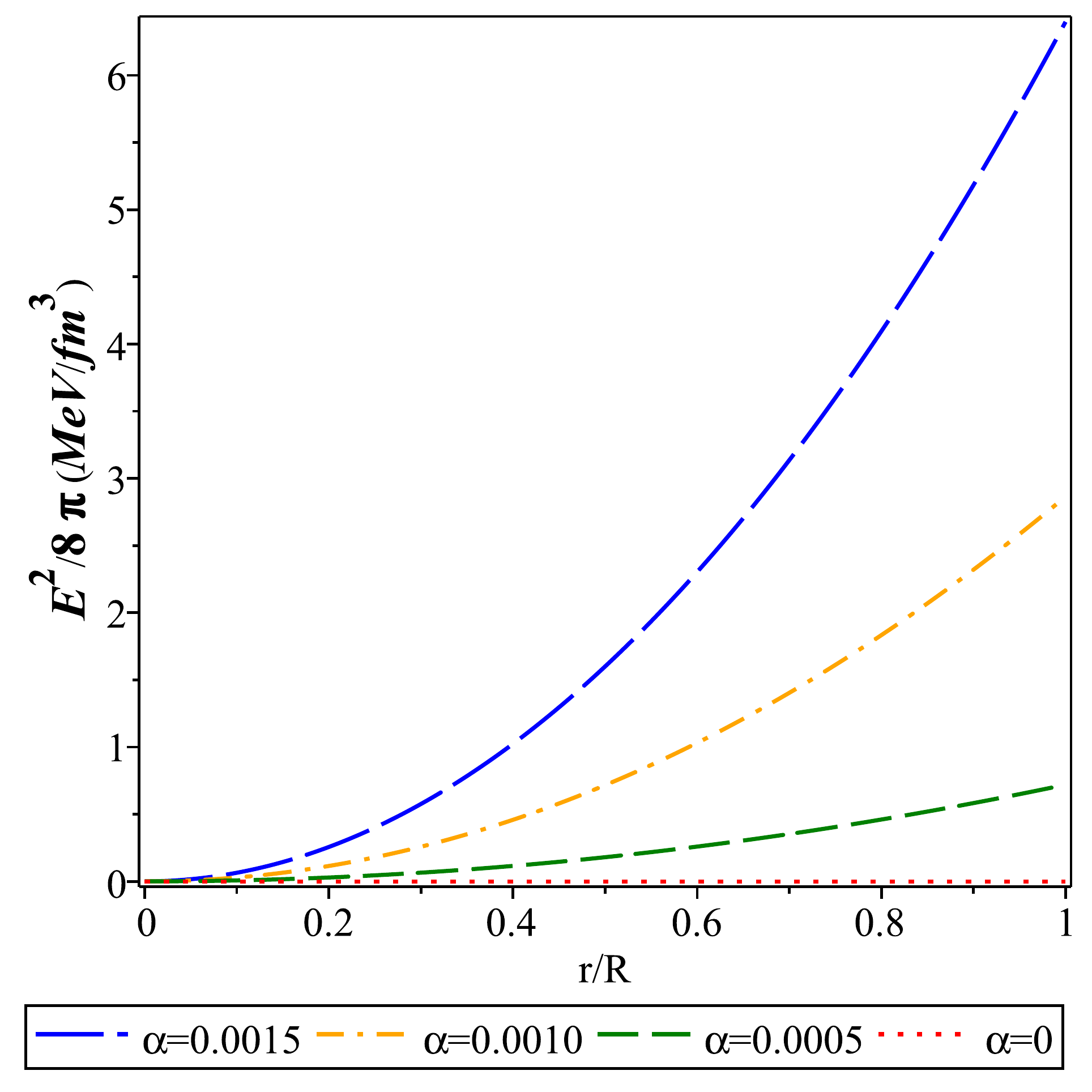}
	\caption{Variation of i) $q(r)$ (left panel) and ii) $ E^2/8\,\pi $ (right panel) as a function of the fractional radial coordinate $r/R$ for the $LMC ~X-4$. } \label{fc}
\end{figure}

The variation of the electrical charge distribution (q(r)) and respective electrical energy density ($ E^2/8\,\pi $) are exhibited in figure \ref{fc} in the left and right panel respectively as a function of fractional radial coordinate. It is clear from the plots that there are no charge density and corresponding  field at the centre of the stellar distribution.

%%%%%%%%%%%%%%%%%%%%%%%%%%%%%%%%%%%%%%%%%%%%%%%%%%%%%%%%%%%%%%%%%%%%%%%%%%%%%%%%%%%%%%%%%%%%%%%%%%%%%%%%

%%%%%%%%%%%%%%%%%%%%%%%%%%%%%%%%%%%%%%%%%%%%%%%%%%%%%%%%%%%%%%%%%%%%%%%%%%%%%%%%%%%%%%%%%%%%%%%%%%%%%%%%

\section{Physical features of the Stellar System} \label{sec:physical}
In this section, we are going to verify whether our model is in the stable equilibrium and physically valid.

\subsection{physically acceptable behaviour}

\subsubsection{Adiabatic Index}
The stiffness of the EOS of a stellar system for a defined density variation can be characterized from the adiabatic index ($\Gamma $). For an infinitesimal radial perturbation, the dynamical stability has been studied by several authors~\cite{Chandrasekhar,Santos,Horvat,Doneva,Silva}. In the following work~\cite{Heintzmann} it is shown that the adiabatic index must be greater than $4/3$ which can be defined as
\begin{eqnarray}
\Gamma = \left( \frac{p_r + \rho }{p_r}\right)  \frac{dp_r}{d \rho}  = \frac{p_r + \rho }{p_r} v_{sr}^2.\nonumber\\
\end{eqnarray}

\begin{figure}[h]\centering
	\includegraphics[scale=0.3]{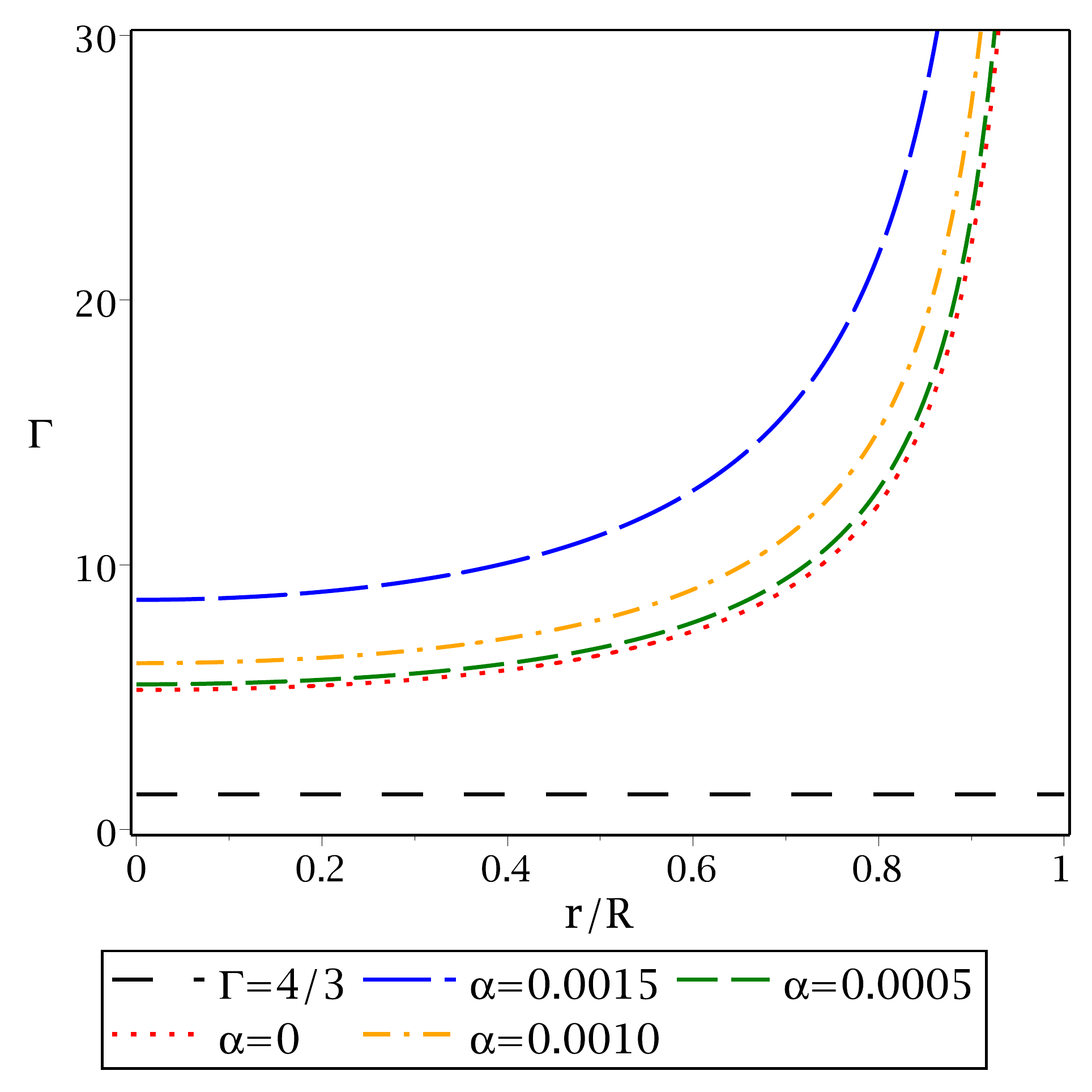}
	\caption{Variation of the adiabatic index as a function of the fractional radial coordinate $r/R$ for the $LMC ~X-4$.} \label{fa}
\end{figure}

The variation of the adiabatic index as a function of the fractional radial coordinate is shown in figure \ref{fa}.

%%%%%%%%%%%%%%%%%%%%%%%%%%%%%%%%%%%%%%%%%%%%%
%%%%%%%%%%%%%%%%%%%%%%%%%%%%%%%%%%%%%%%%%%%%

\subsubsection{Energy Conditions}
Energy conditions depict the observer's measurement of the matter distribution in the space-time. The conditions are always positive, which define that the flow of matter should be along null or time-like world line. Stavrinos and Alexiou~\cite{Stavrinos} provided the required energy conditions for the Finslerian system. A stellar system is said to be a physically valid system if the following inequalities are simultaneously satisfied: 
\begin{eqnarray}
\qquad\hspace{-0.5cm}~NEC&:\rho+p_r\geq 0,~\rho+p_t+\frac{E^2}{4 \pi}\geq 0, \nonumber \\ 
\qquad\hspace{-0.5cm}~WEC&: \rho+p_r\geq 0,~\rho+\frac{E^2}{8 \pi}\geq 0,~\rho+p_t+\frac{E^2}{4 \pi}\geq 0, \nonumber \\
\qquad\hspace{-0.5cm}~SEC&: \rho+p_r\geq 0,~\rho+p_r+2p_t +\frac{E^2}{4 \pi}\geq 0, \nonumber \\ 
\qquad\hspace{-0.5cm}~DEC&:\rho+\frac{E^2}{8 \pi}\geq 0,~ {{\rho}-{p_r}}+\frac{E^2}{4 \pi}\geq 0, ~{{\rho}-{p_t}}\geq 0. \nonumber
\end{eqnarray}

\begin{figure}\centering
	\includegraphics[scale=0.25]{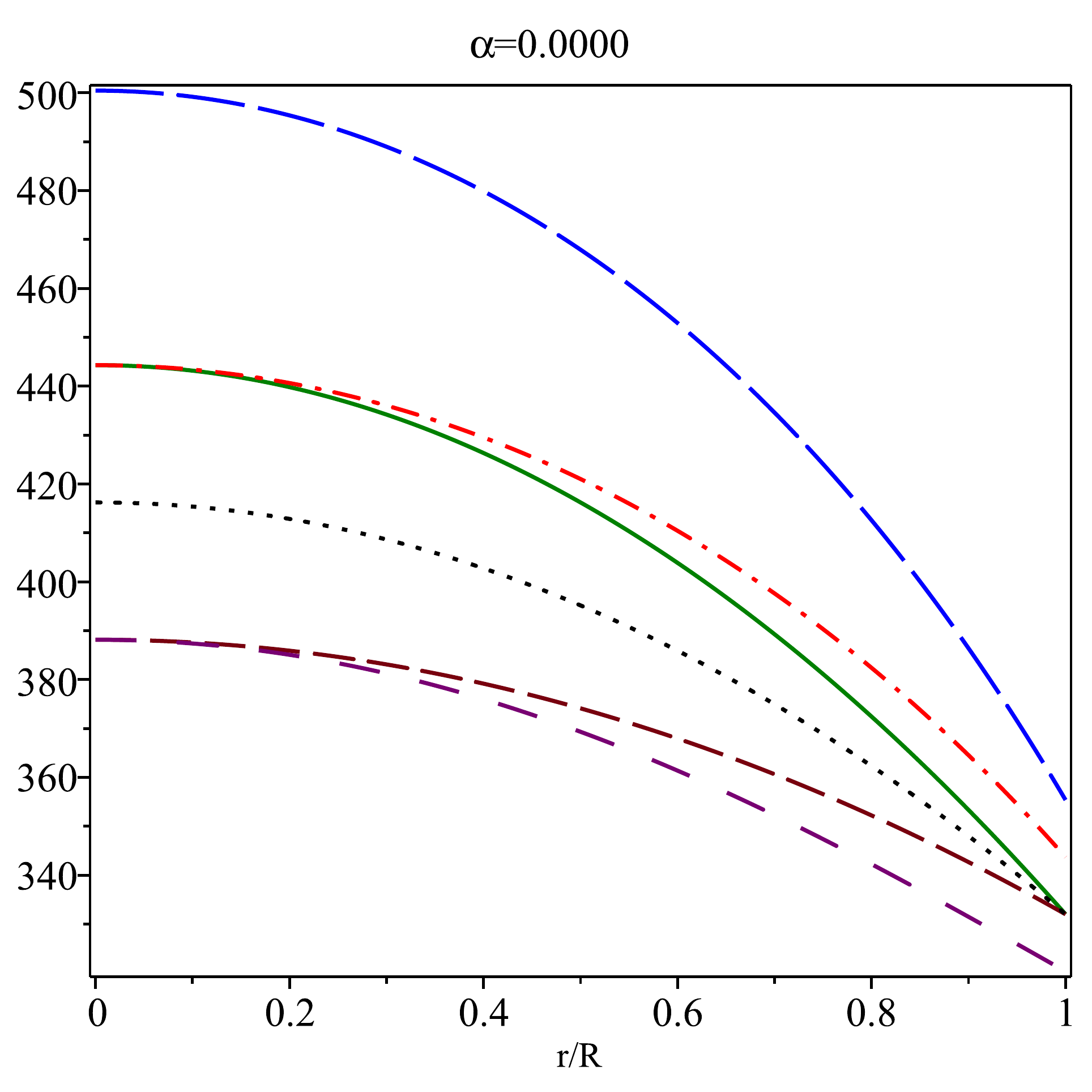}
	~~~\includegraphics[scale=0.25]{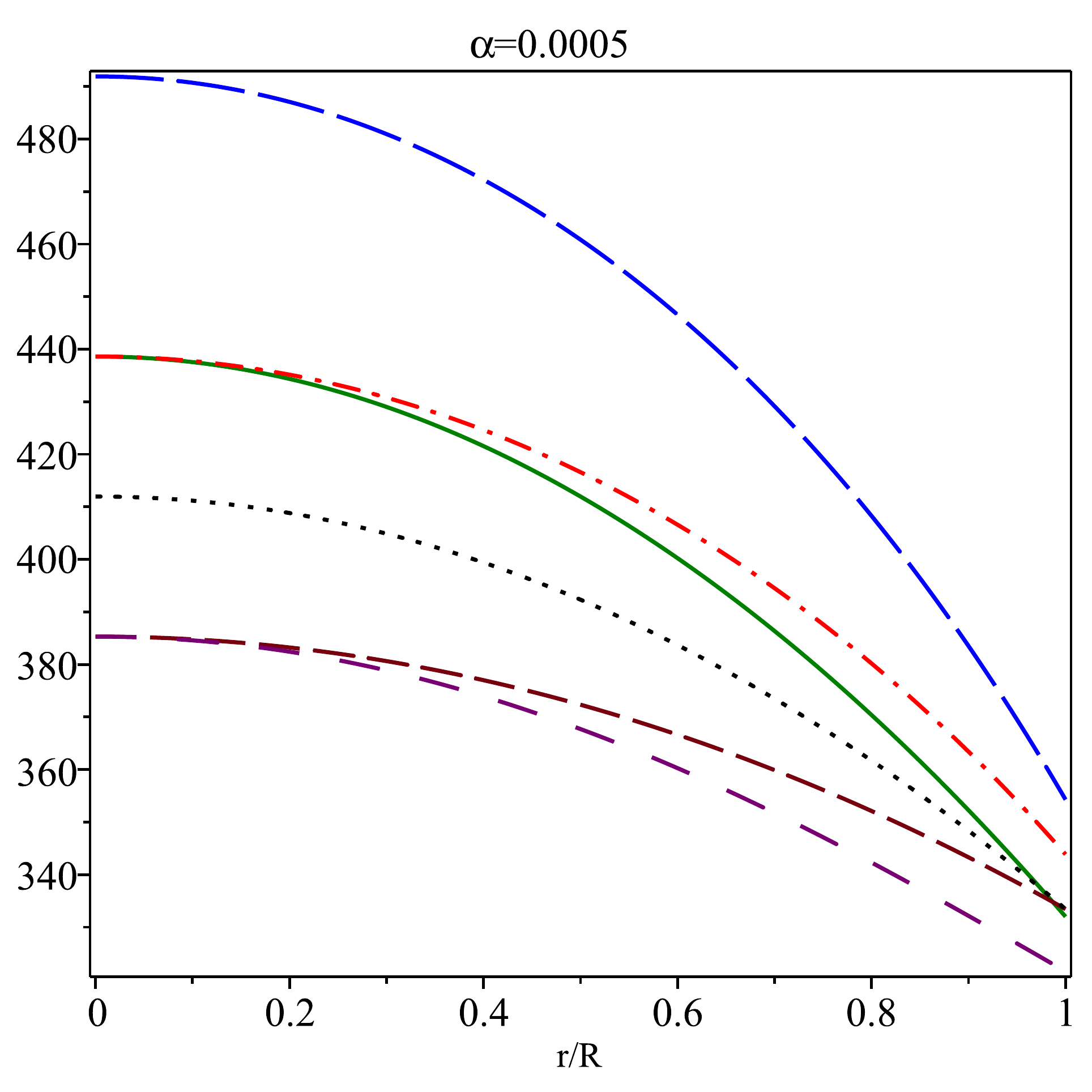}\\
	\includegraphics[scale=0.25]{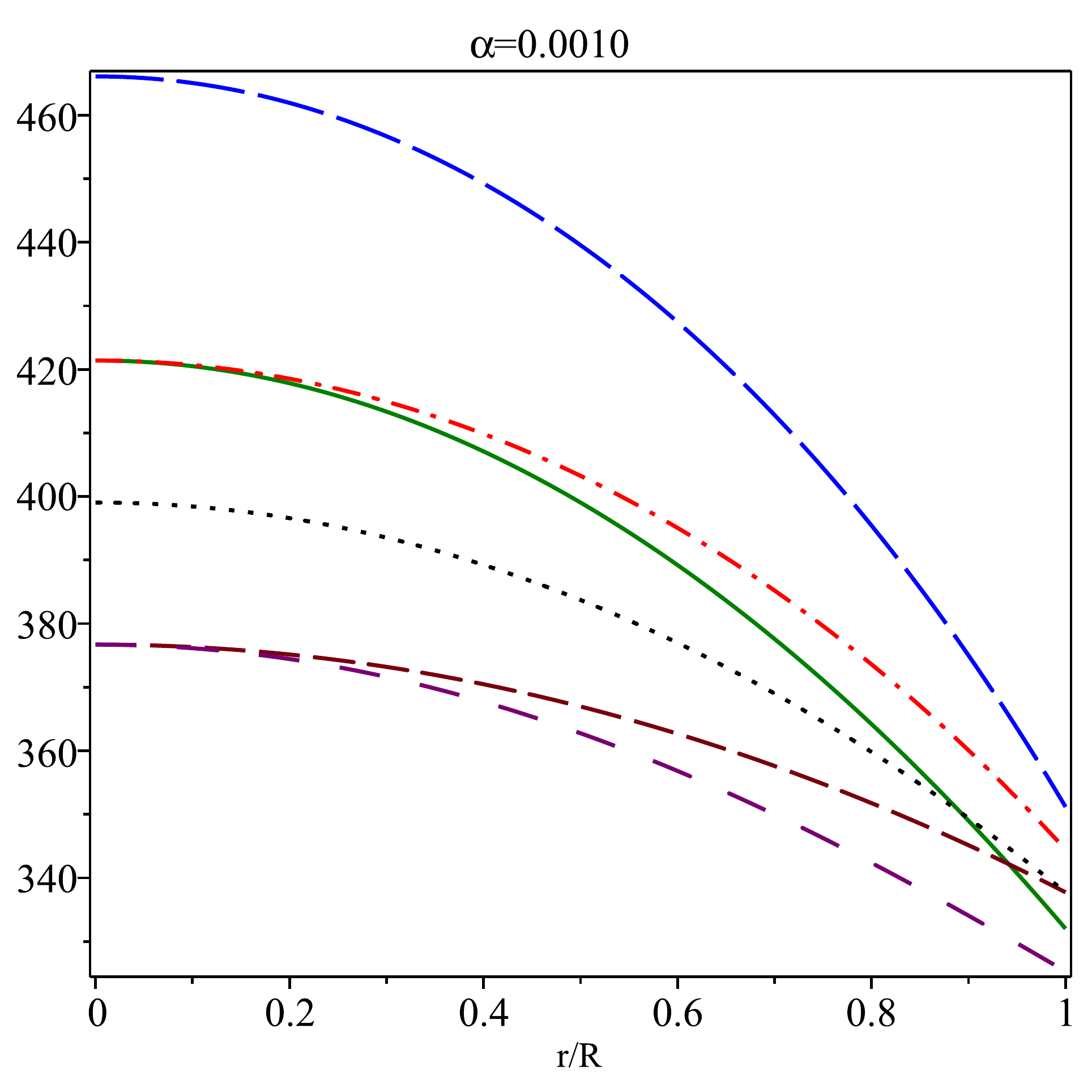}
	~~~\includegraphics[scale=0.25]{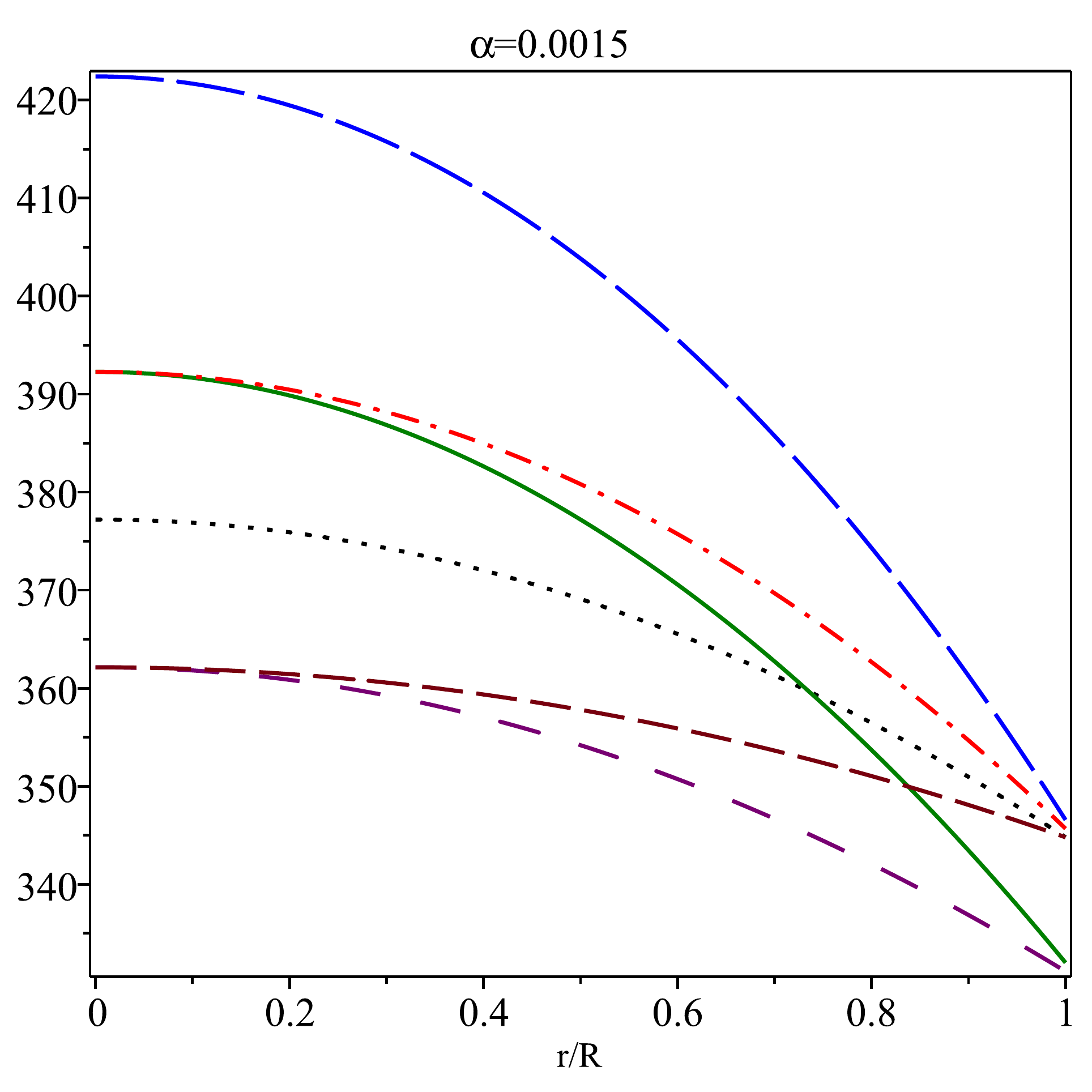}
	\caption{Variation of the different energy conditions as a function of the fractional radial coordinate $r/R$ for the $LMC ~X-4$. Here, blue long dash, red dash dot, green solid, black dot, burgundy dash and purple space dash linestyle stand for $\rho+p_{r}+2\,p_{t}+{\frac {{E}^{2}}{4 \pi}}$, $ \rho+p_t+{\frac {{E}^{2}}{4 \pi}}$, $\rho+p_r$, $\rho+{\frac {{E}^{2}}{ 8 \pi}}$,  $\rho-p_{r}+{\frac {{E}^{2}}{4 \pi}}$ and $ \rho -p_t$, respectively.} \label{fe}
\end{figure}

Here NEC, WEC, SEC and DEC denote the null energy condition, weak energy condition, strong energy condition and dominant energy condition, respectively. The variation of the different energy conditions with the fractional radial coordinate due to different values of $\alpha$ is shown in figure \ref{fe}.

%%%%%%%%%%%%%%%%%%%%%%%%%%%%%%%%%%%%%%%%%%%%%%%%%%
%%%%%%%%%%%%%%%%%%%%%%%%%%%%%%%%%%%%%%%%%%%%%%%%%%

\subsubsection{Mass-Radius Relation}
According to~\cite{Hartle}, the mass-radius ratio for a stable stellar system should maintain $2m(r)/r \leq 1$   throughout the region. Andr{\'e}asson~\cite{Andreasson} generalized the maximum mass-radius ratio for a charged stellar system in the following form
\[
\frac{2M}{R}\leq \frac{2}{9R^2}\left[ 3Q^2 + 2R^2+ 2R \sqrt{3Q^2 + R^2}\right].
\]   

For the present stellar system the mass variation can be written as
\begin{eqnarray}
m \left( r \right) =\frac{{r}^{3}}{2{R}^{5}}\, \big[ -3{R}^{7}{\alpha}^{2
}+3{R}^{5}{\alpha}^{2}{r}^{2}-16B\pi{R}^{5} +16B\pi{R}^{3}{r}^{2}+5M{R}^{2}-3M{r}^{2} \big].
\end{eqnarray}

\begin{figure}\centering
	\includegraphics[scale=0.35]{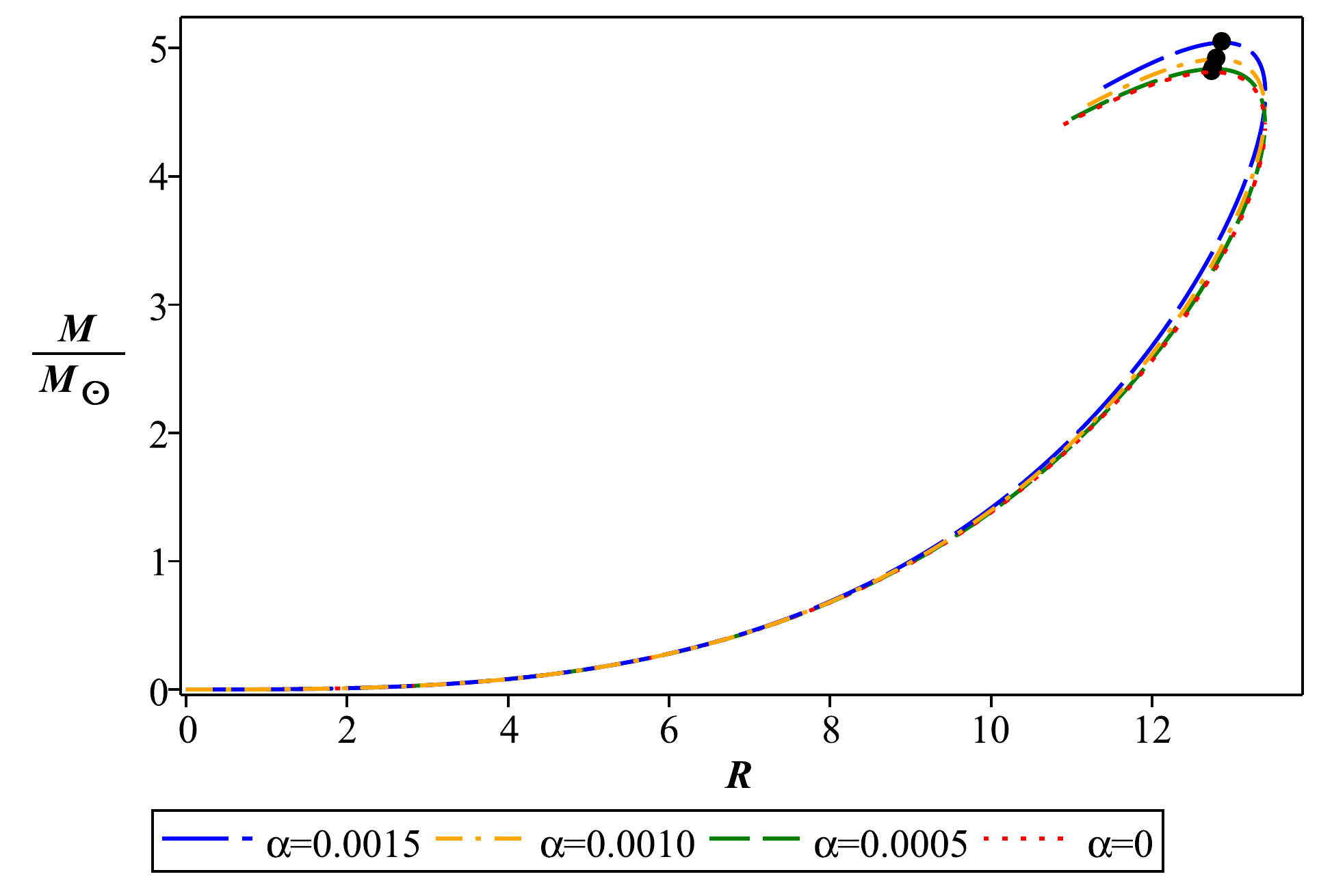}
	\includegraphics[scale=0.35]{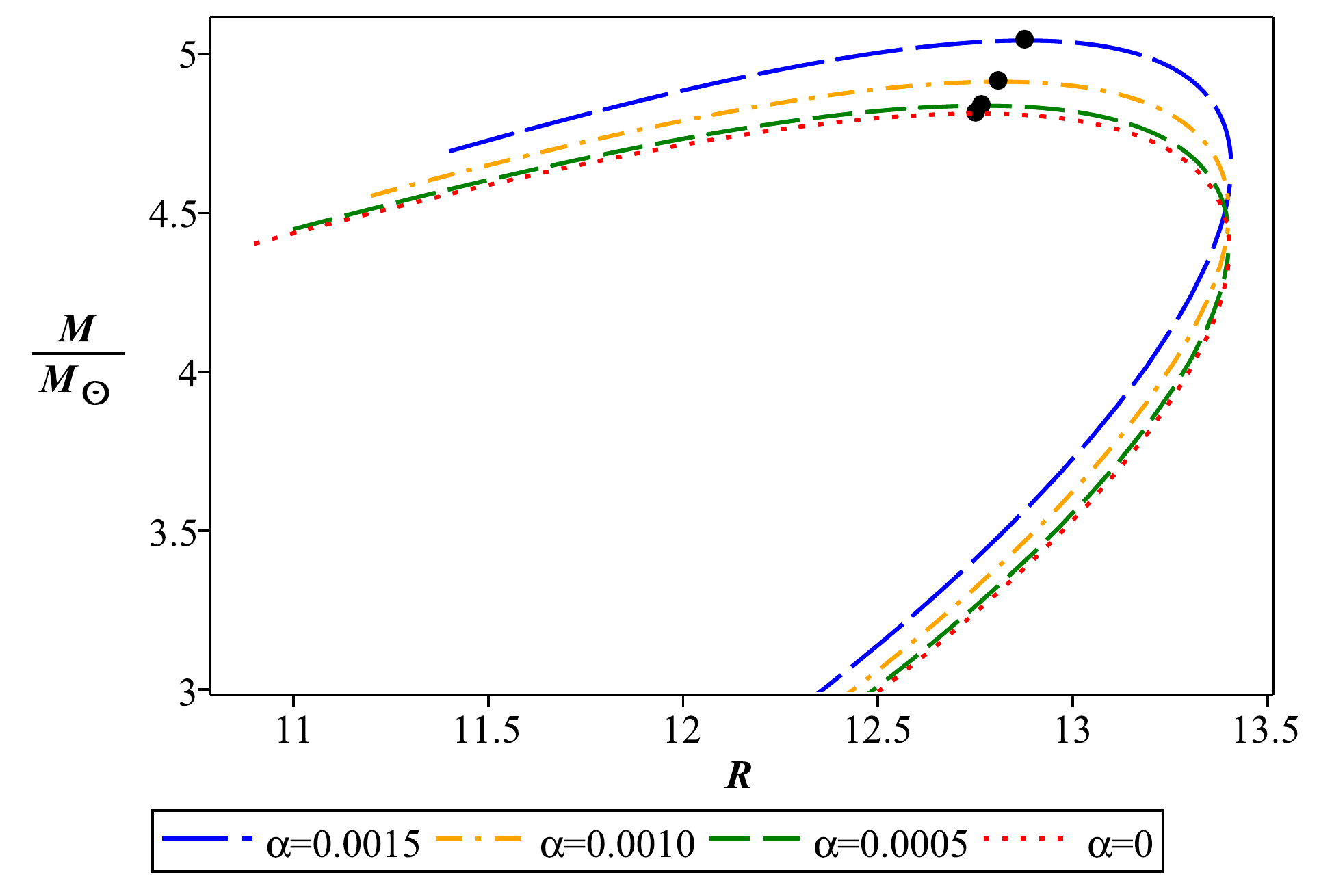}
	\caption{The variation of $M$ (normalized in solar mass $M_{\odot}$) of a strange star as a function of radius shown in the left panel, whereas in the right panel we show enlarged version og $M/M_\odot$ vs $R$ curve. Solid circles are representing the maximum mass and radius of the respective curves.} \label{fmr}
\end{figure}

The variation of total mass respect to total radius is exhibited in figure \ref{fmr}. The variation is drawn for different charge constant. As a result from variation, we obtain that the mass increases with the charge constant. Stellar mass is normalized in solar mass ($M_\odot$).

%%%%%%%%%%%%%%%%%%%%%%%%%%%%%%%%%%%%%%%%%
%%%%%%%%%%%%%%%%%%%%%%%%%%%%%%%%%%%%%%%%%%

\subsubsection{Compactification Factor and Red Shift}

The compactification factor of a stellar system can express as,
\begin{eqnarray}
u(r)= \frac{m(r)}{r}= \frac{{r}^{2}}{2{R}^{5}}\, \Big( -3\,{R}^{7}{\alpha}^{2}+3\,{R}^{5}{\alpha}^{2}{r}^{2}-16\,B\pi\,{R}^{5}+16\,B\pi\,{R}^{3}{r}^{2}\nonumber\\
+5\,M{R}^{2}-3\,M{r}^{2} \Big).
\end{eqnarray} 

The redshift function ($Z_s$) of a stellar system is defined as,
\begin{eqnarray}
& & Z_s = \textrm{e}^{-\lambda /2}-1\nonumber\\
& & \hspace{0.5cm} = \exp \Bigg[\frac {-16}{\lambda_{{1}}\nu_{1}}  \Big[\lambda_{1}\,{\rm arctanh} \left({\frac {{R}^{5}{\alpha}^{2}+16\,B\pi\,{R}^{3}-M}{\lambda_{{2}}}}\right) + \lambda_{{2}} \left( \lambda_{3}-\lambda_{4} \right) \ln \lambda_{5}\nonumber\\
& & \hspace{0.9cm} +\lambda_{1}\,{\rm arctanh} \left({\frac {2 \nu_{1}r^2+2\,\nu_{2}}{\lambda_{{2}}{R}^{2}}}\right)+ \Big\{\lambda_{4}\,\ln  \big(  \left( \nu_{1}r^2+\nu_{2} \right) {r}^{2}\nonumber \\
& & \hspace{0.9cm} +\overline{{\it Ric}} {R}^{5} \big) -\lambda_{3}\,\ln  \overline{{\it Ric}} +\frac{5}{2}\left( {\frac {3\,{R}^{5}{\alpha}^{2}}{40}}+B \pi\,{R}^{3}-\frac{3}{16}\,M \right) \,\ln R   \Big\} \lambda_{{2}}\Big] \Bigg]-1. \nonumber\\ \label{36}
\end{eqnarray}
	
\begin{figure}\centering
	\includegraphics[scale=0.3]{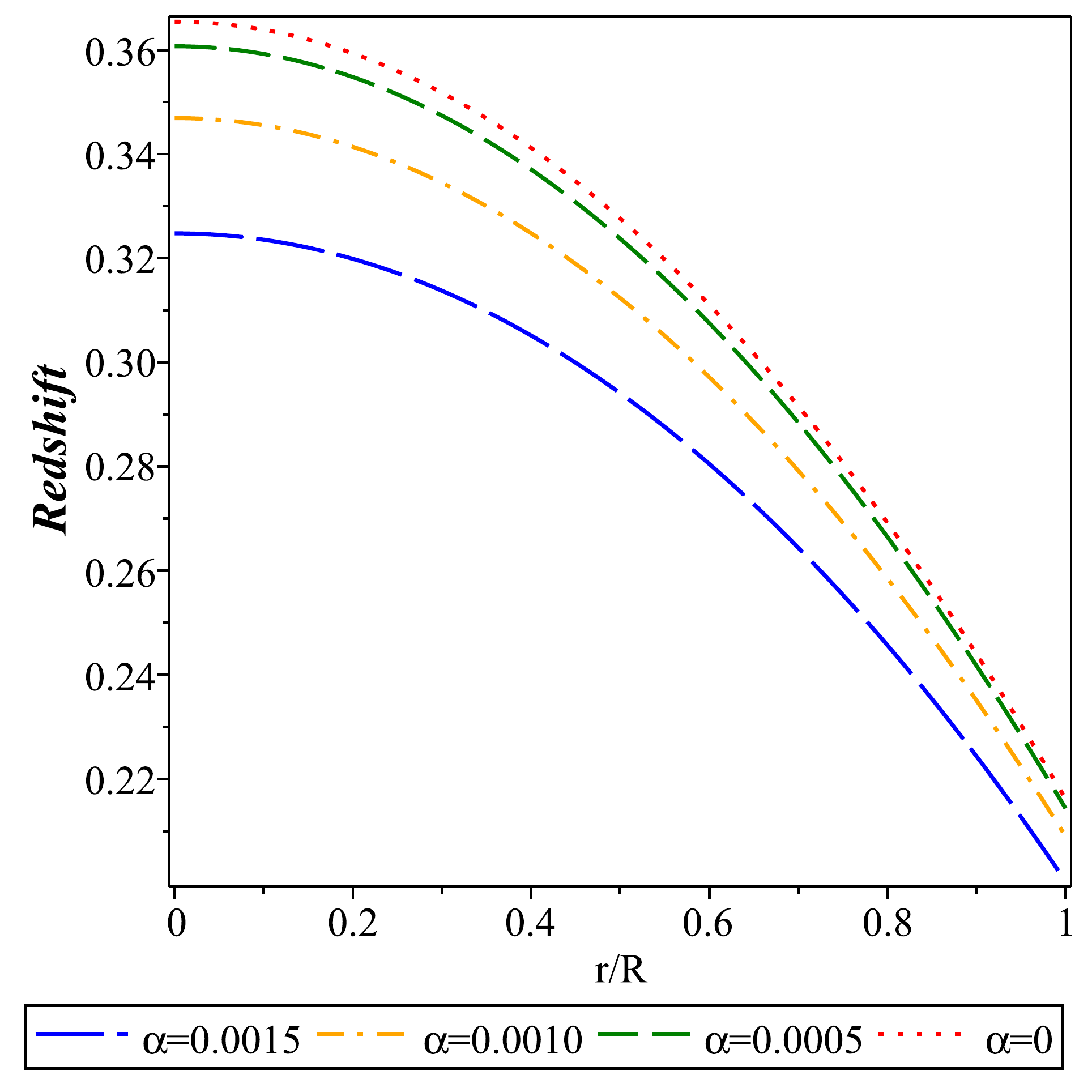}
	\caption{The variation of the redshift function with respect to the fractional radial coordinate for $LMC~X-4$.} \label{fred}
\end{figure}

The variation of $Z_s$ is shown in the figure \ref{fred} as function of radial coordinate. 

%%%%%%%%%%%%%%%%%%%%%%%%%%%%%%%%%%%%%%%%%%%%%%
%%%%%%%%%%%%%%%%%%%%%%%%%%%%%%%%%%%%%%%%%%%%%%

\subsection{Verification of the Stability}
To justify the equilibrium configuration, we considered (i) TOV equation and (ii) Herrera cracking condition.

\subsubsection{TOV Equation}
A stellar system is said to be in equilibrium, if the resultant of forces on a system is nil. Imbalance in force drives the system to an unstable configuration. For the stellar system these counterbalancing forces are given by Tolman~\cite{Tolman} and Oppenheimer-Volkoff~\cite{OV}.

According to TOV equation
\begin{equation}
-p_r'-\frac{\lambda'}{2}(\rho+p_r)+ \sigma\frac{q}{r^2} e^{\nu/2}+\frac{2}{r}(p_t-p_r) = 0. \label{h}
\end{equation}

Here, the terms are defined as hydrostatic force ($  F_{h} $), gravitational force ($  F_{g} $), electrostatic force ($  F_{e} $) and the last term defines the anisotropic force ($  F_{a} $) respectively of the eqs. \ref{h}. To maintain the equilibrium, the outward forces $  F_{e} $, $  F_{a} $ and $  F_{h} $ must balance by the attractive pull $  F_{g} $.

%%%%%%%%%%%%%%%%%%%%%%%%%%%%%%%%%%%%%%%%%%%%
\begin{figure} \centering
	\includegraphics[scale=0.26]{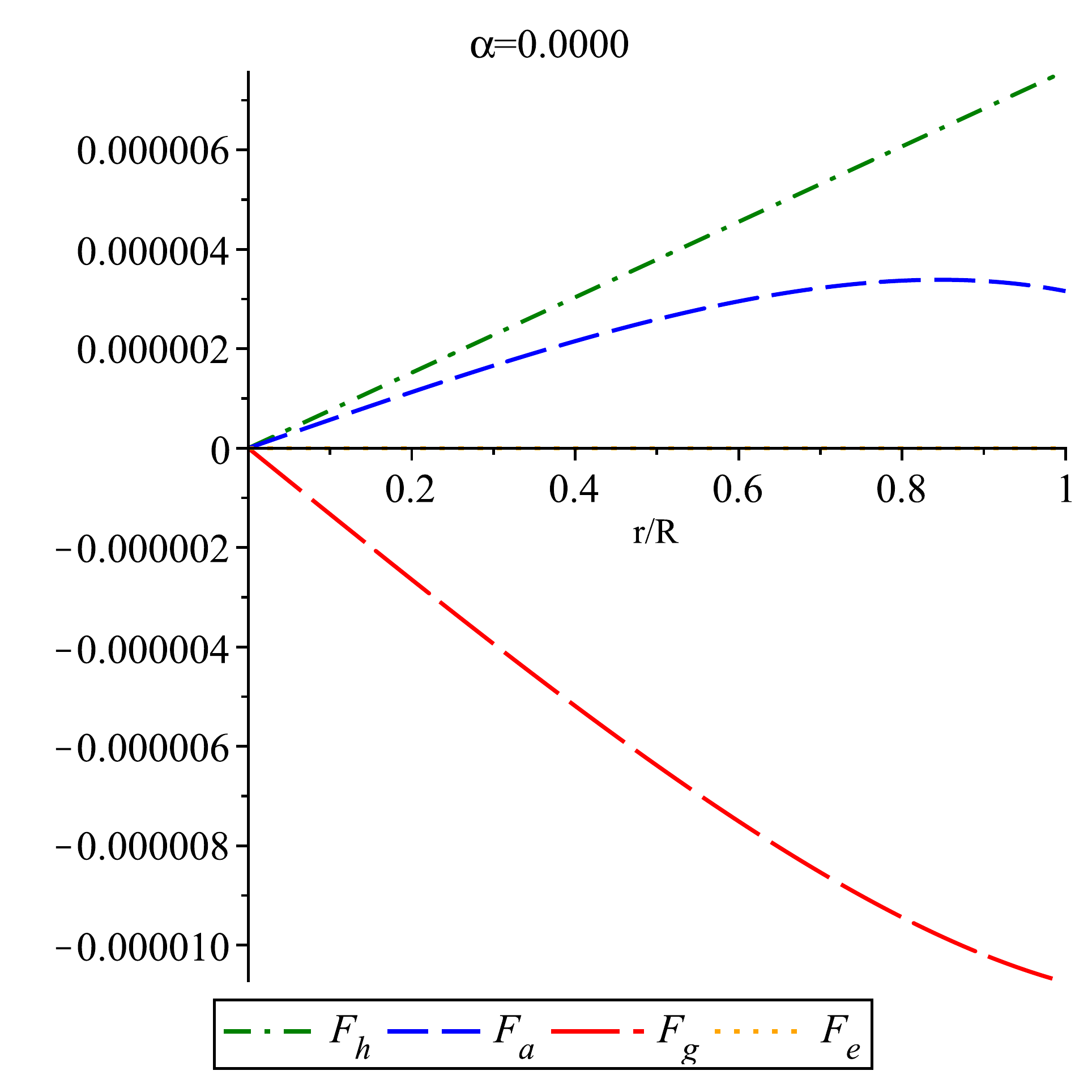}
	~~~\includegraphics[scale=0.26]{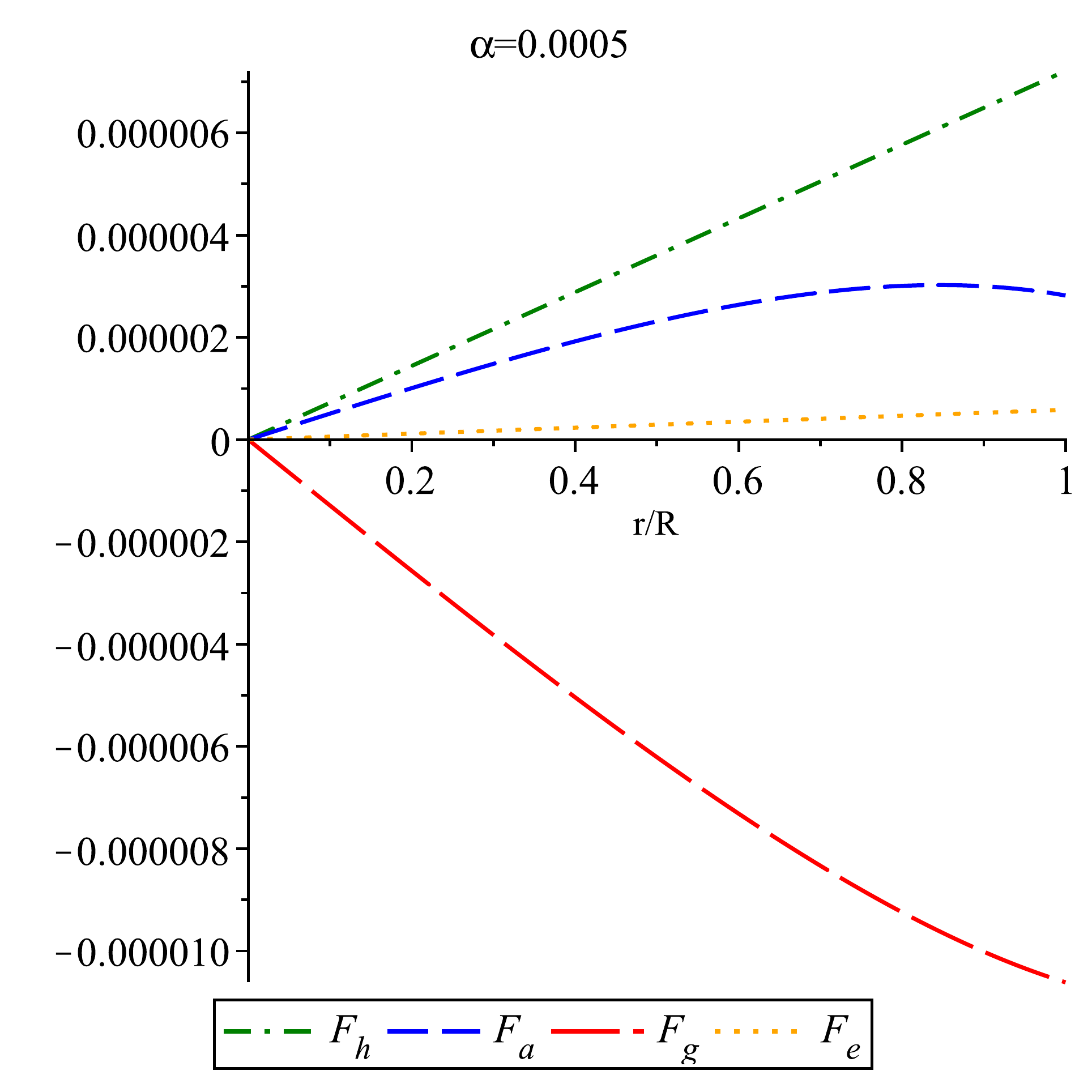}\\
	\includegraphics[scale=0.26]{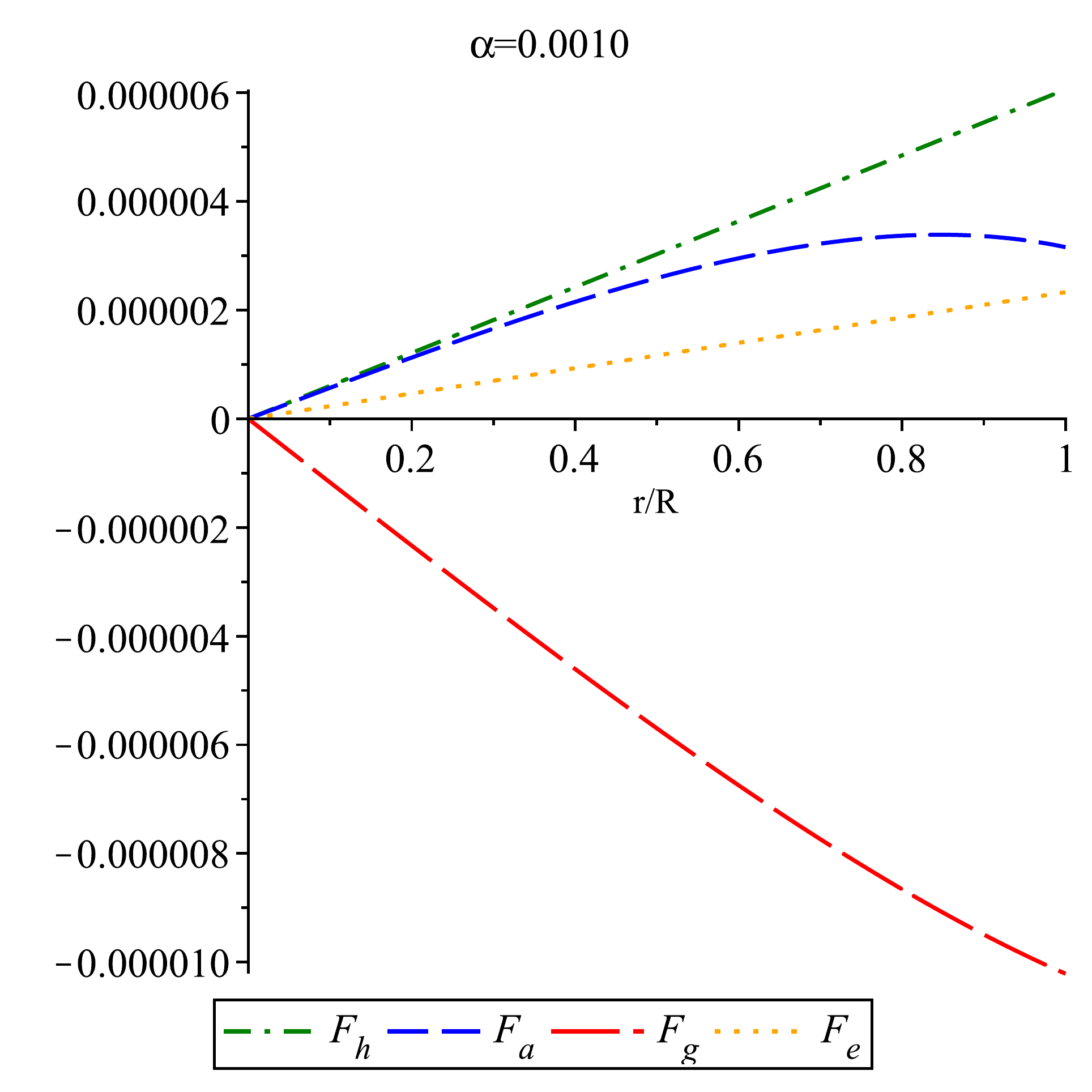}
	~~~\includegraphics[scale=0.26]{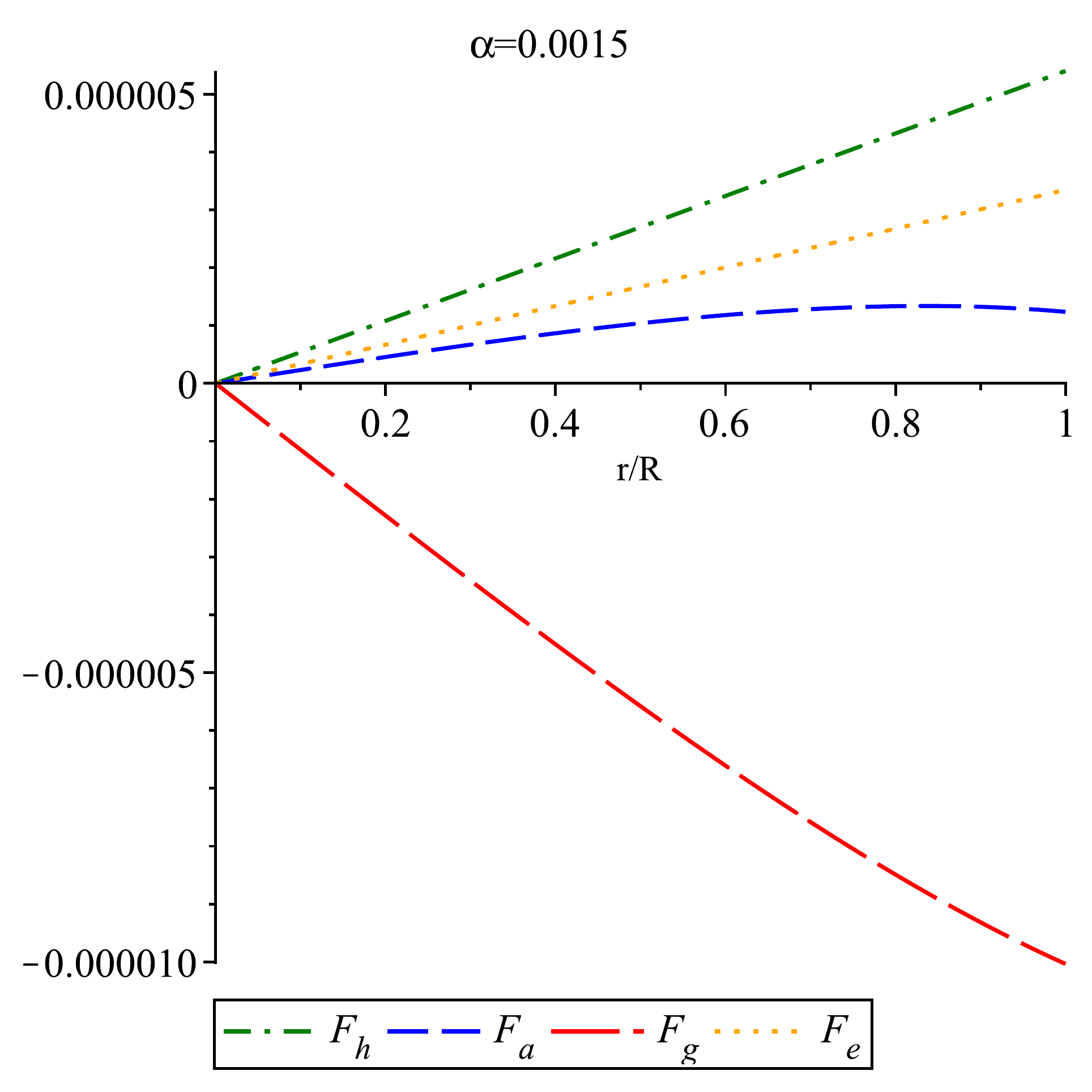}
	\caption{Variation of different forces as a function of the fractional radial coordinate $r/R$ for the $LMC ~X-4$.} \label{fh}
\end{figure}
%%%%%%%%%%%%%%%%%%%%%%%%%%%%%%%%%%%%%%%%%%%%%

The variation of the forces for different charge distributions are exhibit in figure \ref{fh}.

\subsubsection{Herrera cracking Condition}
The physical stability of a stellar system can be verified with respect to the causality condition also. According to the condition, the square of the radial  ($ v_{sr}^2 =\frac{dp_r}{d\rho} $) and the tangential ($ v_{st}^2 =\frac{dp_t}{d\rho}$) sound speed lies between 0 $\rightarrow$ 1, i.e. 0 $\leq v_{si}^2 \leq 1 $ (where, $i=r, t$). The region is called potentially stable if the radial sound speed is greater than the tangential sound speed.

According to the following works~\cite{Herrera,Abreu} the difference of the squares of tangential and radial sound speeds must hold the sign inside the stellar system. Hence, according to Herrera's condition $ \mid v_{st}^2 -v_{sr}^2 \mid \leq $ 1. 

\begin{figure}\centering
	\includegraphics[scale=0.3]{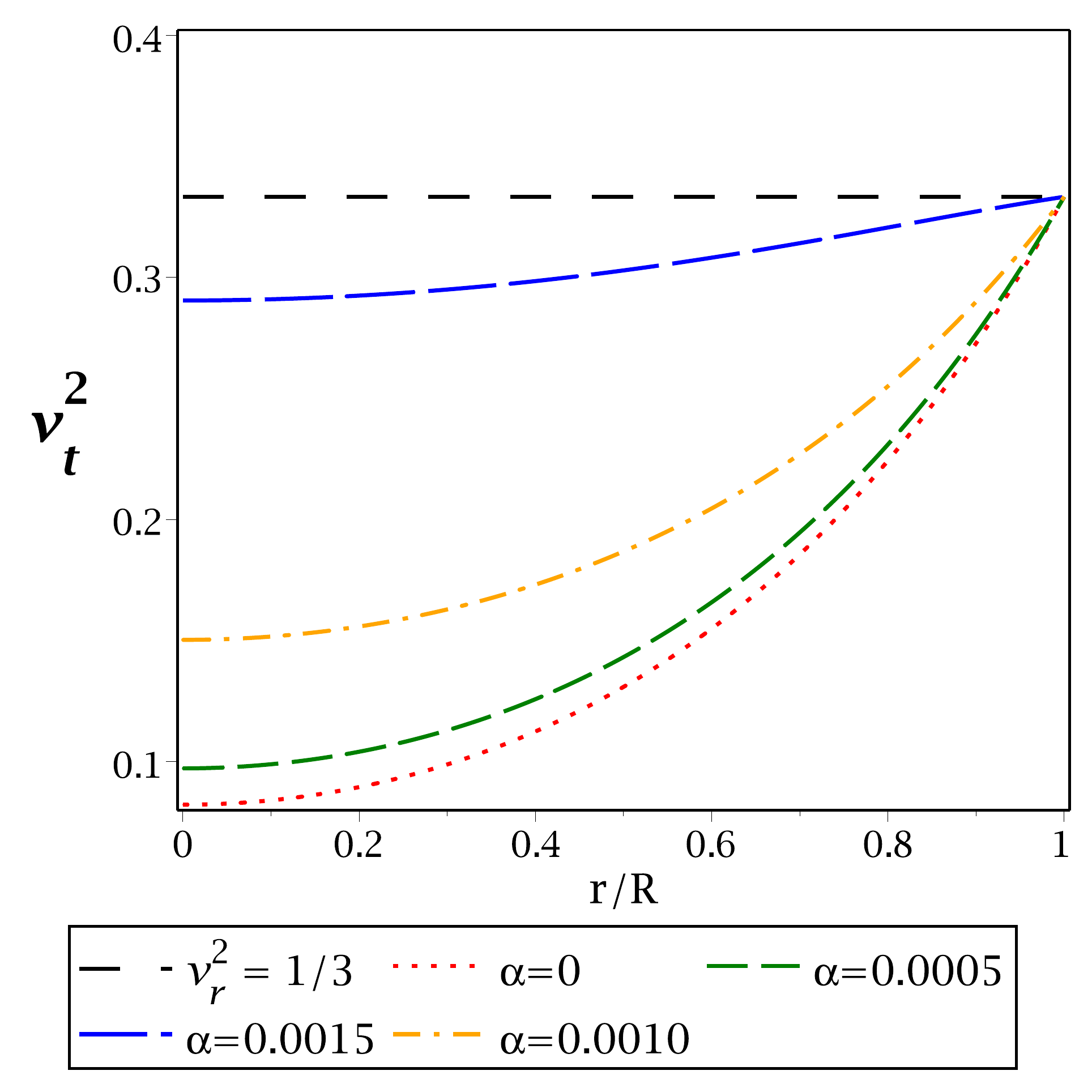}
	\includegraphics[scale=0.3]{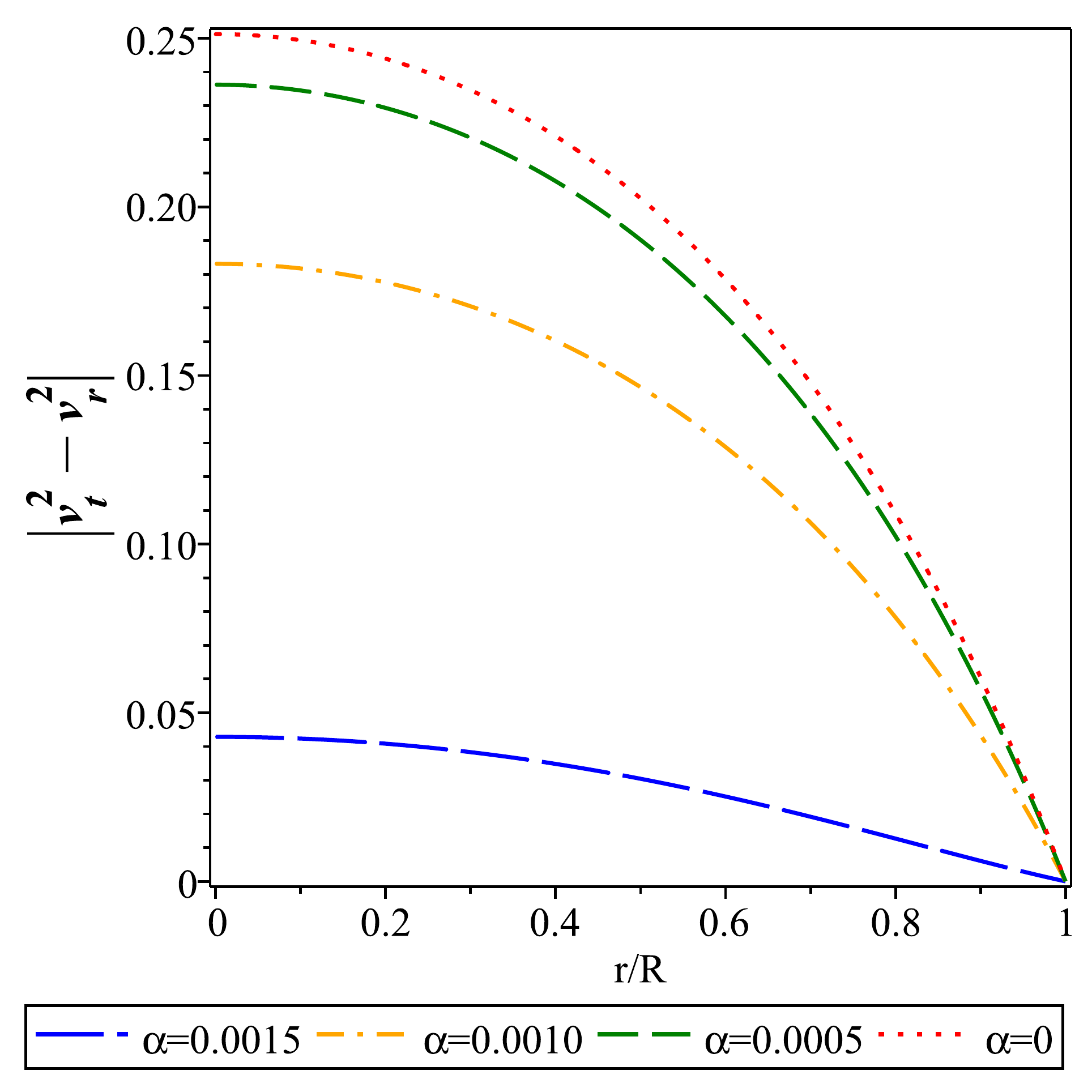}
	\caption{Variation of i) $v_{sr}^2$ and $v_{st}^2$ (left panel) and ii) $ \mid v_{st}^2 -v_{sr}^2 \mid$ (right panel) as a function of the fractional radial coordinate $r/R$ for the $LMC ~X-4$.} \label{fs}
\end{figure}

In figure \ref{fs} the variation of the radial and tangential sound speeds and their differences are shown in the left panel and right panel, respectively.

%%%%%%%%%%%%%%%%%%%%%%%%%%%%%%%%%%%%%%%%%%%%%
%%%%%%%%%%%%%%%%%%%%%%%%%%%%%%%%%%%%%%%%%%%%%

\section{Discussion and conclusion}\label{sec:outlook}
In the paper at hand, we have performed a detail investigation of the physical impacts of the charge distribution on the structure and behaviour of the strange stars. We have generalized the description of the charged strange stars in Finslerian background by studying the modified form of the Maxwell-Einstein field equation. We derived the gravitational potentials from the family of field equations and smoothly matched them with exterior Finslerian Ressiner-Nordstr{\"o}m solution. Figure \ref{fn} indicates the variation of the metric potentials viz., $ e^{\nu(r)} $ and $ e^{\lambda(r)} $, which are monotonically increasing and geometrically non-singular by nature. Interior quarks and electrons distribution profile of the system is related to the radial pressure of the system by the MIT bag equation. The density profile of the system decreases from the central density $\rho_c$~\eqref{cenden} holding the same sign of the slope. The radius of the stellar body is predicted when, there is no overlying matter against the gravitational attractive force, which endorse zero radial pressure. Due to the reason, the surface density is constant, this is one of the limitations of the approach. The radial and tangential pressure also decreases monotonically with the radial variation. The expression and variation of density, radial and tangential pressure are provided in eqs.~\eqref{29}-\eqref{31} and in figure \ref{frho}, respectively. As a consequence of using aniotropic fluid, our model shows that the anisotropic stress is maximum at surface area and there is no anisotropy at the centre of the system, as predicted by Deb et al. \cite{deb} in Riemannian frame. Due to the introduction of charge, the anisotropy reduces with increment of the charge constant, i.e., charge can remove the anisotropy of a system. The expression is provided in eq. (\ref{31}) and the figure \ref{fani}. The variation of charge distribution and the corresponding field is portrayed in figure \ref{fe}. With the considered assumption, the charge and the following electric field reaches its maximum at the surface.  The acceptability of the system is examined on the basis of the energy conditions, Herrera cracking condition, TOV equation and mass-radius relation.

%%%%%%%%%%%%%%%%%%%%%%%%%%%%%%%%%%%%%%%%%%%
\begin{figure}\centering
	\includegraphics[scale=0.3]{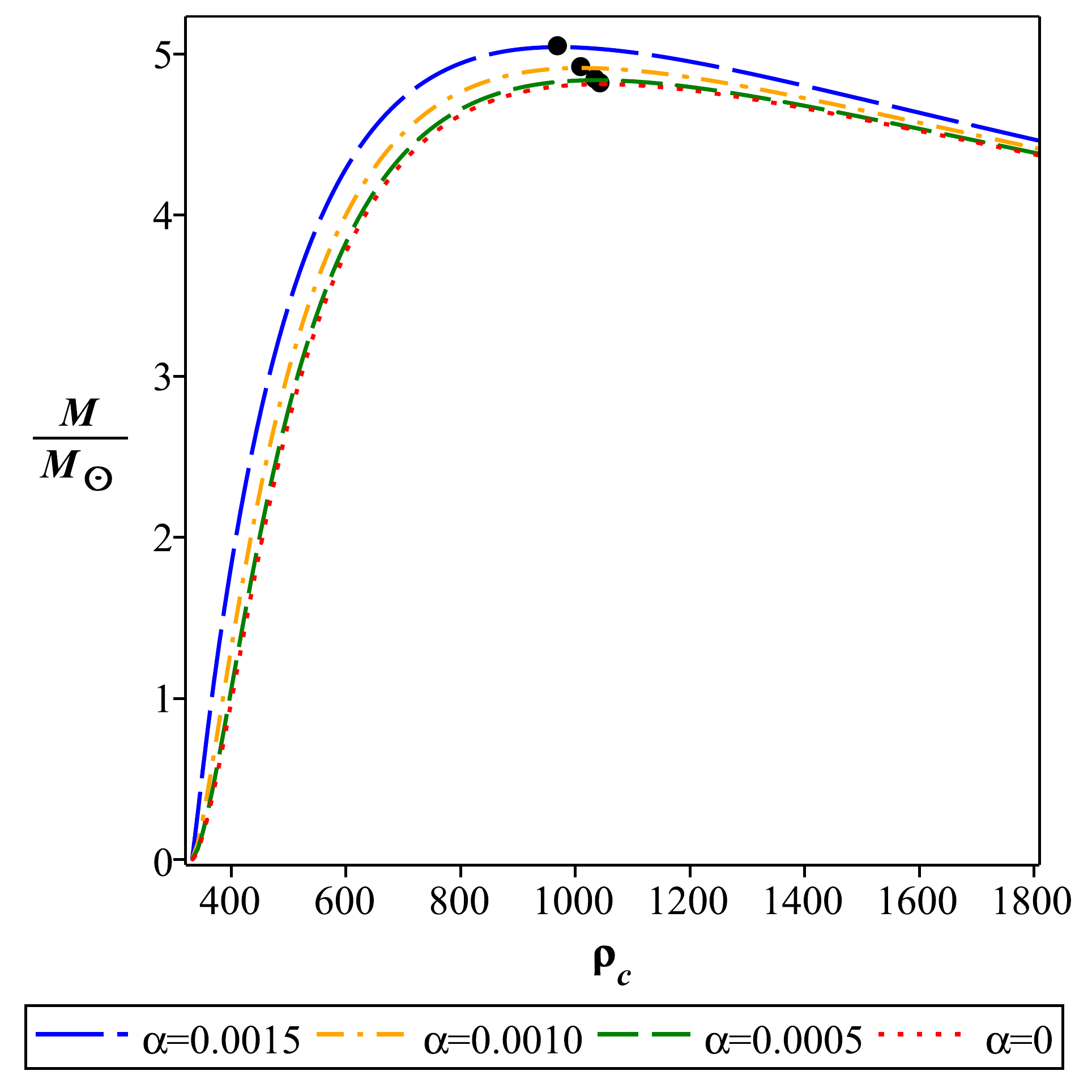}
	\includegraphics[scale=0.3]{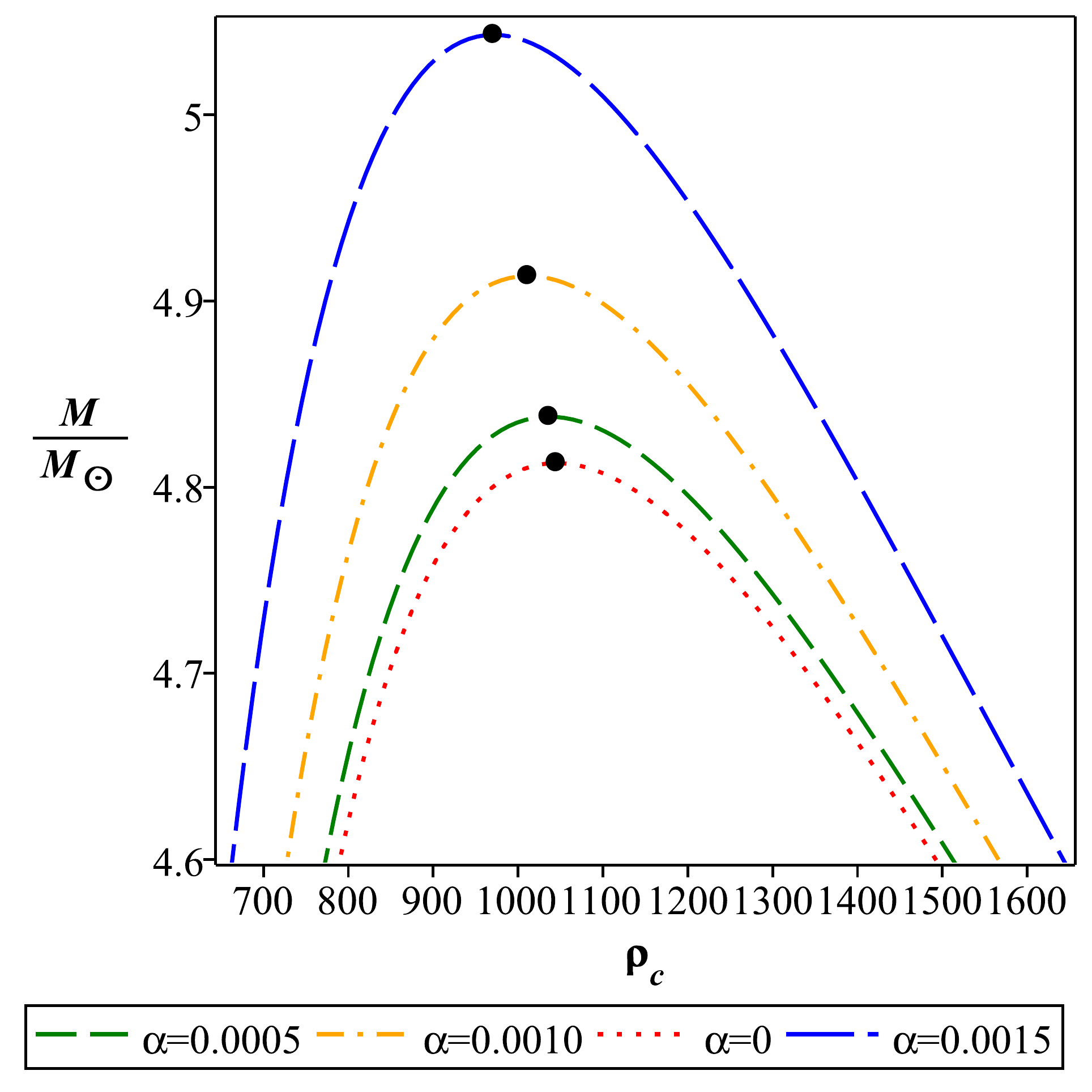} 
	\caption{The variation of $M$ (normalized in $M_{\odot}$) with respect to the central density $({\rho}_c)$ is shown in the left panel, whereas in the right panel we show the enlarged version of $M/M_\odot$ vs ${\rho}_c$ curve. Solid circles are representing the maximum mass for the system.}  \label{fmrho}
\end{figure}
%%%%%%%%%%%%%%%%%%%%%%%%%%%%%%%%%%%%%%%%%%%

%%%%%%%%%%%%%%%%%%%%%%%%%%%%%%%%%%%%%%%%%%%
\begin{figure}\centering
	\includegraphics[scale=0.3]{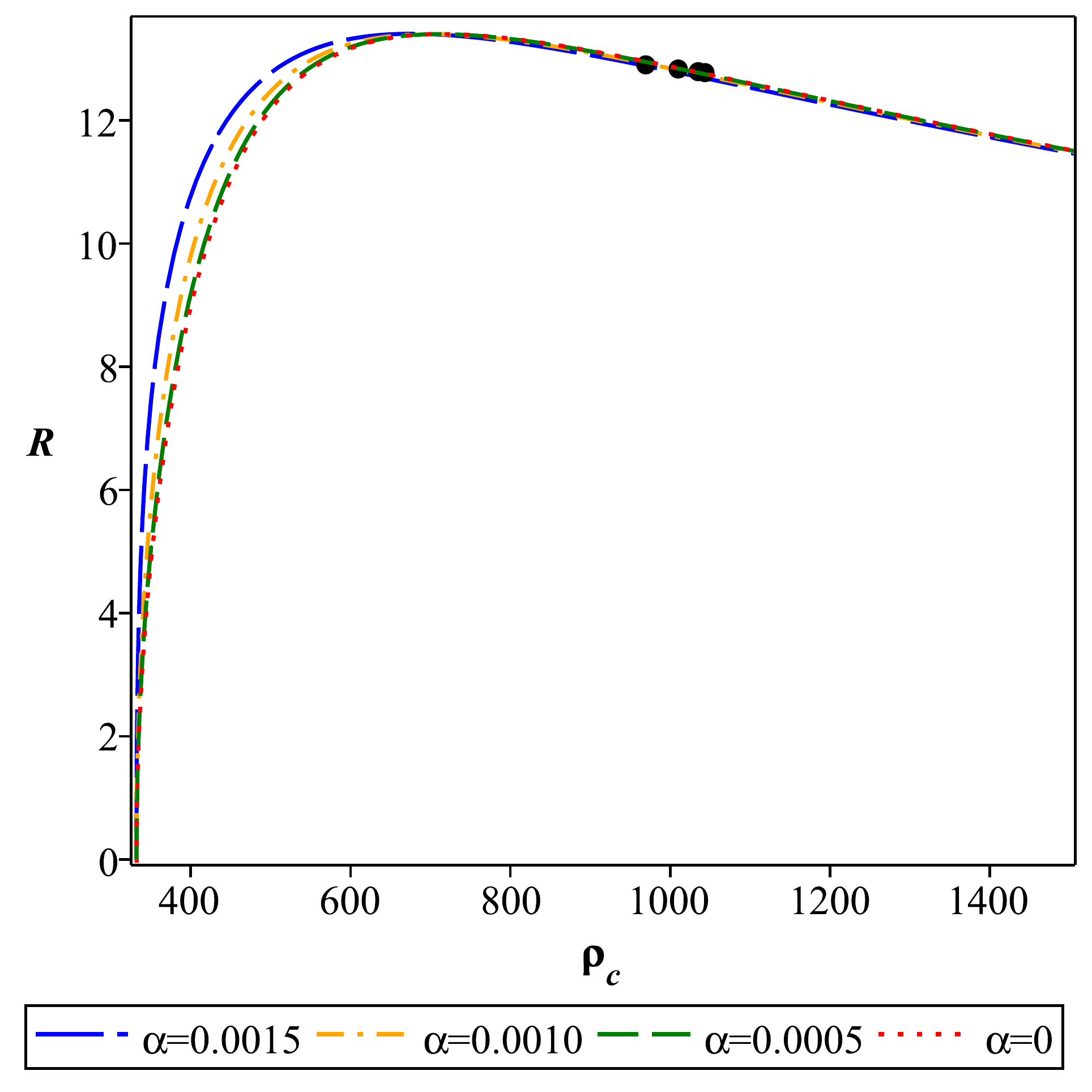}
	\includegraphics[scale=0.3]{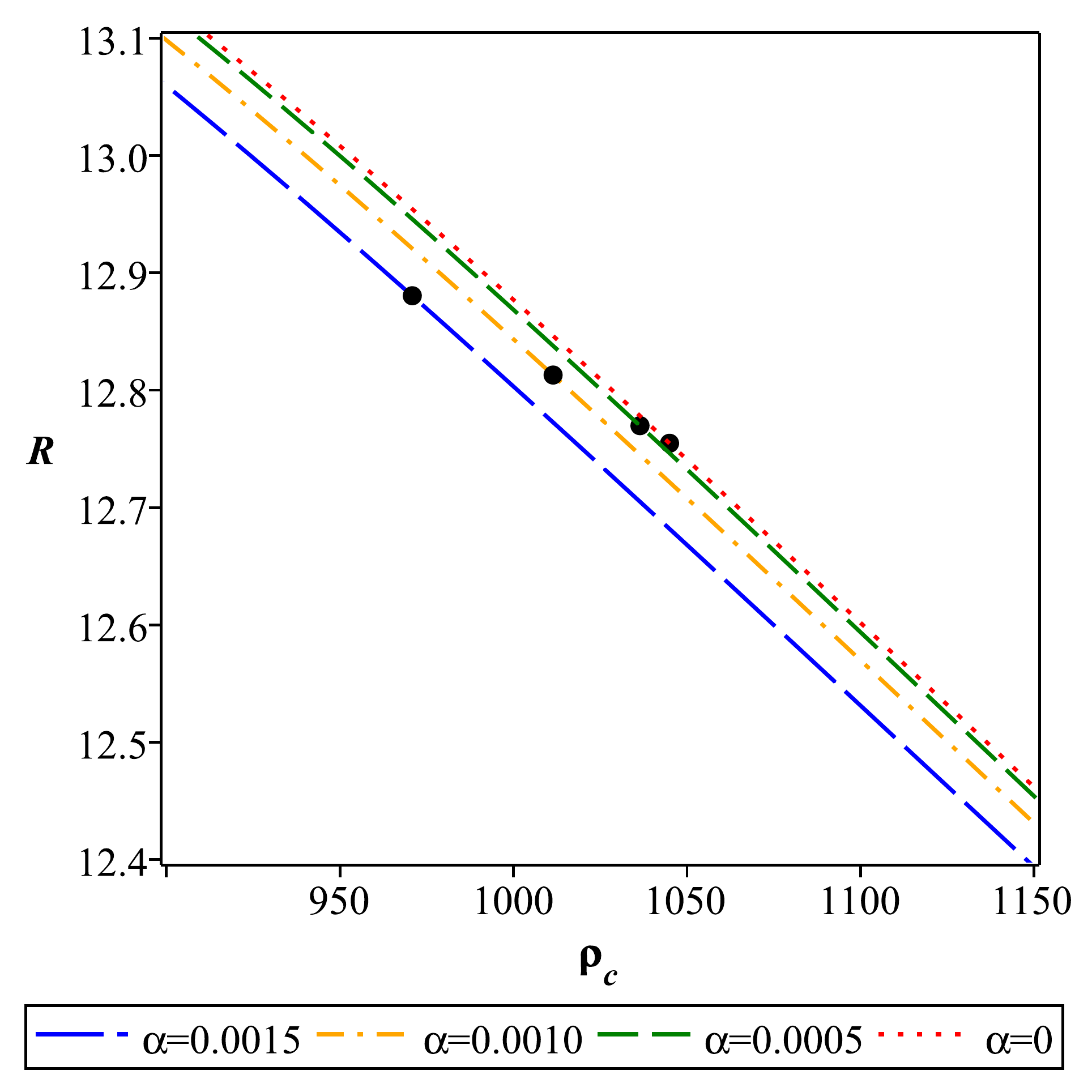}
	\caption{The variation of $R$ with respect to ${\rho}_c$ is shown in the left panel, whereas in the right panel we show the enlarged version of $R$ vs ${\rho}_c$ curve. Solid circles are representing the maximum radius for the system.}  \label{frrho}
\end{figure}
%%%%%%%%%%%%%%%%%%%%%%%%%%%%%%%%%%%%%%%%%%%

%%%%%%%%%%%%%%%%%%%%%%%%%%%%%%%%%%%%%%%%%%%%%%%%%%%%%%%%%%%%%%%%%%%%%%%%%%%%%%%%%%%%%%%%%%%% 

\begin{table}[tbp]
	\centering	
	{\footnotesize
\begin{tabular}{|cccccccc|}
	\hline
			Strange&Observed&Predicted&${\rho}_{c}$&${{P_c}}$&Q&E&Surface  \\
			Stars&Mass&Radius&(${gm/cm}^3$)&(${dyne/cm}^2$)&($ Coulomb$)&(V/m)& redshift  \\ 
			~&$({{M}_{\odot}})$&$km$&$ \times 10^{14} $&$ \times 10^{34} $&$ \times 10^{20} $&$\times 10^{22}$& \\
		\hline
			
			$PSR~J~1614-2230$ & $1.97 \pm 0.04$\cite{demorest} & $11.08^{+0.06}_{-0.07}$  & $7.703  $ & $ 5.314$ & $ 1.578  $ & $1.156 $ &  0.315 \\ 
			
			$Cyg~X-2$ & $1.78 \pm 0.23$\cite{orosz} &  $10.75 ^{+0.39}_{-0.43}$ & $7.555$ & $4.873 $  & $ 1.441  $ & $122 $ & 0.284 \\
			
			$PSR~B1913+16$ & $1.44 \pm 0.002$\cite{van2015} &  $10.09 ^{+0.01}_{-0.01}$ & $7.254 $ & $3.976 $ & $ 1.192 $ & $1.053 $ & 0.231 \\
			
			$Cen~X-3$ & $1.09 \pm 0.08$\cite{van} &  $9.27 ^{+0.20}_{-0.22}$ & $6.933 $ & $3.022 $ & $ 0.924  $ & $0.967$ &  0.180 \\ 
			
			$LMC~X-4$ & $1.29 \pm 0.05$\cite{dey2013} &  $9.76 ^{+0.11}_{-0.12}$ &  $7.116 $ & $3.565$ & $ 1.079 $ & $1.018 $ & 0.209 \\ 
			
			$4U~1820-30$ & $1.58 \pm 0.06$\cite{guver2010b} &  $10.38 ^{+0.12}_{-0.13}$ &  $7.535 $ & $4.276 $ & $ 1.298 $ & $1.083$ & 0.253 \\
			
			$4U~1608-52$ & $1.74 \pm 0.14$\cite{guver2010a} &  $10.68 ^{+0.25}_{-0.26}$ & $7.509 $ & $4.736 $ & $ 1.413 $ & $1.115 $ & 0.278 \\ 
			
			$EXO~1785-248$ & $1.30 \pm 0.20$\cite{ozel2009} &  $9.78 ^{+0.44}_{-0.49}$ & $7.140$ & $3.637 $ & $ 1.085  $ & $1.021$ & 0.253 \\
			
			$Vela~X-1$  & $1.77 \pm 0.16$\cite{Barziv} &  $10.73^{+0.28}_{-0.29}$  & $ 7.535  $ & $4.276 $ & $ 1.433  $ & $1.120 $ & 0.211 \\ 
			\hline
		\end{tabular}
	}
		\caption{\label{tab:table1} The physical parameters are predicted from the proposed model where $1~{{M}_{\odot}}=1.475~km$ for $G=c=1$. The set is valid for B$ _g $ = 83 Mev/fm${^3}$, $ \overline{Ric} $ = 1.2 and $\alpha$ = 0.0010 $km^{-2}$}
	
\end{table}

%%%%%%%%%%%%%%%%%%%%%%%%%%%%%%%%%%%%%%%%%%%%%%%%%%%%%%%%%%%%%%%%%%%%%%%%%%%%%%%%%%%%%%%%%%%% 

%%%%%%%%%%%%%%%%%%%%%%%%%%%%%%%%%%%%%%%%%%%%%%
\begin{figure}\centering
	\includegraphics[scale=0.3]{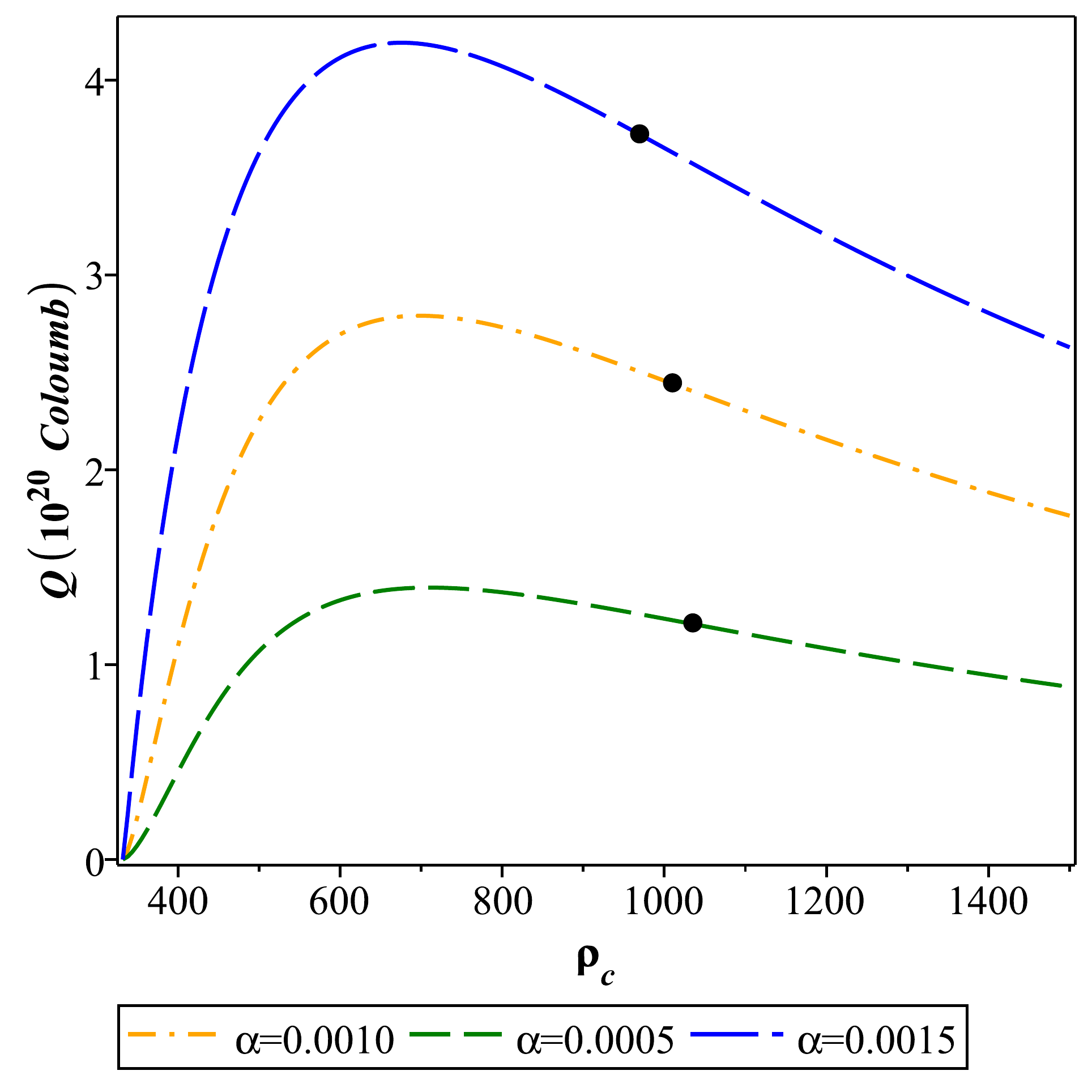}
	\caption{The variation of the total charge ($Q$) with respect to ${\rho}_c$ is shown in the left panel, whereas in the right panel we show the enlarged version of $Q$ vs ${\rho}_c$ curve. Solid circles are representing the maximum charge for the system.}  \label{fqrho}
\end{figure}
%%%%%%%%%%%%%%%%%%%%%%%%%%%%%%%%%%%%%%%%%%%%%%

%%%%%%%%%%%%%%%%%%%%%%%%%%%%%%%%%%%%%%%%%%%%%%%%%%%%%%%%%%%%%%%%%%%%%%%%%%%%%%%%%%%%%%%%%%%% 

\begin{table}[tbp]
	\centering 
\begin{tabular}{|ccccc|}
	\hline
			
			Value of $ \alpha $   & $ \alpha $ = 0.0000  &  $ \alpha $ = 0.0005&  $ \alpha $ = 0.0010 &  $ \alpha $ = 0.0015 \\
			 &$ km^{-2} $ &$ km^{-2} $ &$ km^{-2} $ &$ km^{-2} $\\  \hline 
			
			Predicted Radius ($km$) &  9.79 &  9.78 &   9.76 &   9.72  \\
			
			Central density $({gm/cm}^3)$ &  $7.421\times 10^{14} $ &   $7.359\times 10^{14} $ &  $7.116\times 10^{14} $ &   $6.754\times 10^{14} $ \\ 
			
			Central pressure $({dyne/cm}^2)$ &  $4.484\times 10^{34} $ &  $4.290\times 10^{34} $ &  $3.565\times 10^{34} $ &  $2.488\times 10^{34} $\\ 
			
			Charge $(Coulomb)$  &  $ 0.000 $ &  $0.543\times 10^{20} $ &  $1.079\times 10^{20} $ &  $1.598\times 10^{20} $ \\ 
			
			Electric field ($V/m$) &  $ 0.000 $ &  $0.510\times 10^{22} $ &  $1.016\times 10^{22} $ &  $1.521\times 10^{22} $\\
			
			Surface redshift ($Z_s$) &  0.216 &  0.214 &  0.209 &  0.201\\ 
	\hline		
		\end{tabular}
	\caption{\label{tab:table2} Numerical values of physical parameters for different charge constant {$\alpha$} for the strange star $LMC~X-4$ of mass 1.29$M_\odot$ ($1~{{M}_{\odot}}=1.475~km$) with B$ _g $ = 83 Mev/fm${^3}$. 
	}
\end{table}

%%%%%%%%%%%%%%%%%%%%%%%%%%%%%%%%%%%%%%%%%%%%%%%%%%%%%%%%%%%%%%%%%%%%%%%%%%%%%%%%%%%%%%%%%%%% 

The anisotropic flow along with the pressure gradient and coulomb repulsion is supported by gravitational pull of the matter (mass), inwards to it. It defines the overlying matter density reduces radially. The variation of the forces endorses that our stellar structure is non-varying in terms of the equilibrium of forces.

Cracking concept describes the nature of the deviation of the system from equilibrium. The notion depends on the theory of gravitation not on the geometry. We found that our stellar structure is consistent with both conditions (i) the causality relation and (ii) the Herrera cracking concept. 

The red shift is monotonically decreasing with radial variation toward the surface. Variation of the redshift as a function of fractional radial coordinate ($r/R$) for the strange star $LMC~X-4$ is portrait in figure \ref{fred}, where the compactification factor defines the mass boundness of the structure.

%%%%%%%%%%%%%%%%%%%%%%%%%%%%%%%%%%%%%%%%%%%%%%%%%%%%%%%%%%%%%%%%%%%%%%%%%%%%%%%%%%%%%%%%%%%% 

\begin{table}[tbp]
	\centering 
	\begin{tabular}{|cccc|}
	\hline
			
			Value of $ \overline{Ric} $  &$ \overline{Ric} $ = 1 & $ \overline{Ric} $ = 1.1 & $ \overline{Ric} $ = 1.2 \\ 
			\hline 
			
			Predicted Radius ($km$) &  9.68 &  9.73 &   9.76\\
			
			Central density $({gm/cm}^3)$ &  $7.531\times 10^{14} $ &   $7.372\times 10^{14} $ &  $7.116\times 10^{14} $ \\ 
			
			Central pressure $({dyne/cm}^2)$ &  $4.812\times 10^{34} $ &  $4.028\times 10^{34} $ &  $3.565\times 10^{34} $\\ 
			
			Charge $(Coulomb)$ &  $1.052\times 10^{20} $ &  $1.068\times 10^{20} $ &  $1.079\times 10^{20} $ \\ 
			
			Electric field ($V/m$) &  $ 1.010 \times 10^{22} $ &  $1.015\times 10^{22} $ &  $1.018\times 10^{22} $ \\
			
			Surface redshif  &  0.274 &  0.237 &  0.209 \\  		
	\hline			
		\end{tabular}
	\caption{\label{tab:table3} Numerical values of physical parameters for different {$\overline{Ric}$} for the strange star $LMC~X-4$ of mass 1.29$M_\odot$ ($1~{{M}_{\odot}}=1.475~km$) with B$ _g $ = 83 Mev/fm${^3}$ and $\alpha$ = 0.0010 $km^{-2}$.}
\end{table}

%%%%%%%%%%%%%%%%%%%%%%%%%%%%%%%%%%%%%%%%%%%%%%%%%%%%%%%%%%%%%%%%%%%%%%%%%%%%%%%%%%%%%%%%%%%% 

%%%%%%%%%%%%%%%%%%%%%%%%%%%%%%%%%%%%%%%%%%%%%%%%%%%%%%%%%%%%%%%%%%%%%%%%%%%%%%%%%%%%%%%%%%%% 

\begin{table}[tbp]
	\centering
\begin{tabular}{|cccc|}
	\hline
			Value of B$ _g $  & B$ _g $ = 70 Mev/fm$ ^3 $ & B$ _g $ = 80 Mev/fm$ ^3 $ &  B$ _g $ = 90 Mev/fm$ ^3 $ \\ \hline
			
			Predicted Radius($km$) &  10.33 &  9.88 &   9.50  \\
			
			Central density $({gm/cm}^3)$ &  $5.875\times 10^{14} $ &   $6.833\times 10^{14} $  &  $7.779\times 10^{14} $ \\ 
			
			Central pressure $({dyne/cm}^2)$ &  $2.631\times 10^{34} $ &   $3.361\times 10^{34} $  &  $4.055\times 10^{34} $ \\ 
			
			Charge $(Coulomb)$ &  $1.279\times 10^{20} $ &  $1.119\times 10^{20} $ &  $1.0.995\times 10^{20} $ \\ 
			
			Electric field ($V/m$) &  $ 1.078 \times 10^{22} $ &  $1.031\times 10^{22} $ &  $0.992\times 10^{22} $ \\
			
			Surface redshift  &  0.192 &  0.205 &  0.217 \\ 
	\hline
\end{tabular}
\caption{\label{tab:table4} Numerical values of physical parameters for different Bag values for the strange star $LMC~X-4$ of mass 1.29$M_\odot$ ($1~{{M}_{\odot}}=1.475~km$) with $ \overline{Ric} $ = 1.2.}
\end{table}

%%%%%%%%%%%%%%%%%%%%%%%%%%%%%%%%%%%%%%%%%%%%%%%%%%%%%%%%%%%%%%%%%%%%%%%%%%%%%%%%%%%%%%%%%%%% 

We have predicted the radii of few candidates of strange stars by using their observed masses. In addition we studied different physical parameters related to the structure for the predicted radius. All studies are provided in the tabular form, Ref. Table \ref{tab:table1}. 

The maximum limiting mass and correlated radius gradually increase with the increasing values of $ \alpha $ ref. figure \ref{fmr} (the solid circles in the variation of the total mass M of the the stellar structure corresponds to maximum mass limit of the model for a given value of  $ \overline{Ric} =1.2 $ and $ B_g=83\,MeV/fm{^3} $). We found that the maximum mass ($M_{max} $) for $ \alpha = 0.0015 ~km^{-2}$  is higher than $M _{max} $ for zero charge by $4.77 \%$ and the respective radius $R _{max} $ is increased by 0.98\%. A stellar configuration is a stable or unstable equilibrium, can  be categorized from the relation $\frac{dM}{d \rho_c} $. Positive slope indicates stable configuration. Maximum mass point for $ \alpha = 0.0015~ km^{-2}$ is increased by $7.6\%$ than the maximum mass point for zero charge respective of $\rho_c$ variation. The variations of $M$ (normalized in $M_\odot$) and $R$ with respect to $\rho_c$ are shown in the figures \ref{fmrho} and \ref{frrho}, respectively. In addition, the variation of total charge ($Q$) with respect to central density is shown in figure~\ref{fqrho}, which features that with the increasing values of $Q$ the central density of the strange stars decreases gradually. Again, for the charge constant $\alpha = 0.0015~ km^{-2}$ the maximum total charge is obtained at $\rho_c=963~ Mev/fm^3$, whereas the maximum mass $M_{max}=5.03 M_\odot$ is obtained for $\rho_c=965~ Mev/fm^3$. From the above numerical study, it is interesting to mention that, in the presence of charge, the amount of total mass increases more intensely than the radius, which indicates that the more compact systems can be justified in the framework. 

 To have a better understanding of the present model in Tables \ref{tab:table2} and \ref{tab:table3} we present a comparative study of the different physical parameters, viz., radius, central density, central pressure, surface redshift, total charge of the structure and the corresponding field, etc., due to the chosen parametric values of $\alpha$ and  $ \overline{Ric}$, respectively. Further, we have predicted values of the above mentioned physical parameters due to the chosen values of bag constant, viz. $B=70, 80$ and $90~MeV/fm^3$ in Table~\ref {tab:table4}. Interestingly, the numerical analysis reveals that with the increasing values of $ \overline{Ric} $ it is possible to pack more mass in the stellar system and the total charge of the system also increases gradually. 

There are also few limitations in our approach, due to which simulation is not completely analogues with the real life system, formally applicable to define a static stellar structure. Within all constraints, it is worthy of noting that in Finsler geometry the ultra-high dense strange stars can be successfully represented, which leaves the very interesting area for future research.

\section*{Acknowledgments}
SR and FR are thankful to the Inter-University Centre
for Astronomy and Astrophysics (IUCAA), Pune, India
for providing Visiting Associateship under which a
part of this work was carried out. SR is also thankful to
the authority of The Institute of Mathematical Sciences,
Chennai, India and the Centre for Theoretical Studies,
IIT Kharagpur, India for providing short term visits under
which a part of this work was carried out. FR is
also grateful to DST-SERB (EMR/2016/000193), Government
of India for providing financial support. A part
of this work was completed while SRC and DD were visiting
IUCAA and the authors gratefully acknowledge the
warm hospitality and facilities there.

\appendix

\section{Appendix}

Values of the different constants viz., $\nu_1, \nu_2,\lambda_{1},\lambda_{2},\lambda_{3},\lambda_{4}$ and $\lambda_{5}$ are given by
\begin{eqnarray}
& & \nu_1= -2\,{R}^{5}{\alpha}^{2}-16\,B\pi\,{R}^{3}+3\,M, \\
& & \nu_2= 3\,{R}^{7}{\alpha}^{2}+16\,B\pi\,{R}^{5}-5\,M{R}^{2},\\
& & \lambda_{1}={\frac {9\,{R}^{10}{\alpha}^{4}}{16}}+{\frac {15\,{M}^{2}}{16}}+8\,B\pi\,{R}^{8}{\alpha}^{2}-11\,BM\pi\,{R}^{3}  +32\,{B}^{2}{\pi}^{2}{R}^{6}-\frac{3}{2}\,M{R}^{5}{\alpha}^{2}, \\
& & \lambda_{{2}}=\Big(9\,{R}^{10}{\alpha}^{4}+96\,B\pi\,{R}^{8}{\alpha}^{2}+256\,{B}^{2}{\pi}^{2}{R}^{6}+8\,{R}^{6}{\alpha}^{2}\overline{{\it Ric}} -30\,M{R}^{5}{\alpha}^{2}\nonumber \\
& & \hspace{3cm}+64\,B\pi\,{R}^{4}\overline{{\it Ric}} -160\,BM\pi\,{R}^{3}-12\,MR\overline{{\it Ric}} +25
\,{M}^{2}\Big)^{\frac{1}{2}}, \\
& & \lambda_3 = -\frac{9}{32} M+ \frac{3}{2} B \pi R^3+\frac{3}{16} R^5 \alpha^2,\\
& & \lambda_{4}= \frac{3}{16}\,M-B\pi\,{R}^{3}-{\frac {3}{32}\,{R}^{5}{\alpha}^{2}},\\
& & \lambda_{5}={R}^{5}{\alpha}^{2}+R\overline{{\it Ric}}-2\,M,\\ \label{cenden}
& & \rho_{{c}}={\frac {1}{8\,\pi {R}^{3}}\left( -9\,{R}^{5}{\alpha}^{2}-48\,B\pi\,{R}^{3}+15\,M \right) }.
\end{eqnarray}

\end{document}